\documentstyle[12pt]{article}
\@addtoreset{equation}{section}
\def\theequation{\arabic{section}.\arabic{equation}}

\def\section{\@startsection{section}{1}{\z@}{3.5ex plus 1ex minus
   .2ex}{2.3ex plus .2ex}{\large\bf}}
\def\thesection{\Roman{section}.}

\def\appendix{\setcounter{section}{0}
        \def\thesection{Appendix\ \Alph{section}.}
        \def\theequation{\Alph{section}.\arabic{equation}}}

\newcommand{\beq}{\begin{eqnarray}}
\newcommand{\eeq}{\end{eqnarray}}
\newcommand{\eq}{eqnarray}

\newcommand{\al}{{\alpha}}
\newcommand{\be}{{\beta}}
\newcommand{\ci}{\cite}
\newcommand{\ga}{{\gamma}}

\newcommand{\ep}{{\epsilon}}

\newcommand{\de}{{\delta}}
\newcommand{\De}{\Delta}

\newcommand{\La}{{\Lambda}}
\newcommand{\m}{{\mu}}
\newcommand{\n}{{\nu}}

\newcommand{\Si}{{\Sigma}}
\newcommand{\om}{{\omega}}
\newcommand{\Om}{{\Omega}}
\newcommand{\pa}{{\partial}}
\newcommand{\no}{{\nonumber}}
\newcommand{\f}{\frac}
\newcommand{\ra}{\rightarrow}
\newcommand{\lra}{\leftrightarrow}
\newcommand{\eff}{\hbox{\scriptsize eff}}
\newcommand{\tot}{\hbox{\scriptsize tot}}
\newcommand{\stat}{\hbox{\scriptsize stat}}
\newcommand{\exact}{\hbox{\scriptsize exact}}

\newcommand{\new}{\hbox{\scriptsize new}}

\newcommand{\hb}{\hat{\be}}
\newcommand{\we}{\wedge}
\newcommand{\ther}{thermodynamics }

\begin{document}
\topmargin 0pt \oddsidemargin -3.5mm \headheight 0pt \topskip 0mm
\addtolength{\baselineskip}{0.20\baselineskip}
\begin{flushright}
   hep-th/0608165
\end{flushright}
\vspace{0.1cm}
\begin{center}
  {\large \bf BTZ Black Hole with Gravitational Chern-Simons: Thermodynamics and
Statistical Entropy}
\end{center}
\vspace{0.1cm}
\begin{center}
 Mu-In Park\footnote{Electronic address:
muinpark@yahoo.com}
\\{Center for Quantum Spacetime,
}\\{Sogang University, Seoul 121-742, Korea\footnote{Present
address: Research Institute of Physics and Chemistry, Chonbuk
National University, Chonju 561-756, Korea}}
\end{center}
\vspace{0.1cm}
\begin{center}
  {\bf ABSTRACT}
\end{center}

  Recently, the BTZ black hole in the presence of the gravitational
Chern-Simons term has been studied and it is found that the usual
thermodynamic quantities, like as the black hole mass, angular
momentum, and entropy, are modified. But, for large values of the
gravitational Chern-Simons coupling where the modification terms
dominate the original terms some exotic behaviors occur, like as the
roles of the mass and angular momentum are interchanged and the
entropy depends more on the $inner$-horizon area than the outer one.
A basic physical problem of this system is that the form of entropy
does not guarantee the second law of thermodynamics, in contrast to
the Bekenstein-Hawking entropy. Moreover, this entropy does $not$
agree with the statistical entropy, in contrast to a good agreement
for small values of the gravitational Chern-Simons coupling. Here I
find that there is another entropy formula where the usual
Bekenstein-Hawking form dominates the inner-horizon term again, as
in the small gravitational Chern-Simons coupling case, such as the
second law of thermodynamics can be guaranteed.
I also find that the new entropy formula
agrees with the statistical entropy based on the holographic
anomalies for the $whole$ range of the gravitational Chern-Simons
coupling. This reproduces, in the limit of vanishing
Einstein-Hilbert term, the recent result about the exotic BTZ black
holes, where their masses and angular momenta are completely
interchanged and the entropies depend only on the area of the
$inner$ horizon. I compare the result of the holographic approach
with the classical-symmetry-algebra-based approach, and I find exact
agreements  even with the higher-derivative term of gravitational
Chern-Simons. This provides a non-trivial check of the
AdS/CFT-correspondence, in the presence of higher-derivative terms
in the gravity action.

\vspace{0.1cm}
\begin{flushleft}
28 Nov. 2007 \\
\end{flushleft}
\newpage

\begin{section}
{Introduction}
\end{section}

The gravitational Chern-Simons term in Einstein gravity, with a
$vanishing$ cosmological constant $\La$, produces a propagating,
massive, spin-2 mode, although the separate actions do not
\cite{Dese:82,Dese:91}.
( This system is known as the
``Topologically Massive Gravity'' (TMG)  in the literatures. ) So, a
massive object in this theory has the gravitational Chern-Simons
$dressing$ whose size is governed by the inverse of the graviton's
mass, which is proportional to the coupling constant of the
gravitational Chern-Simons term.

Recently, the BTZ black hole system as a $trivial$ solution of the
gravitational-Chern-Simons-corrected gravity in three-dimensional
anti-de Sitter space (AdS) with a negative cosmological constant
$\Lambda=-1/l^2$ ( I call this the {\it
gravitational-Chern-Simons-corrected/dressed BTZ (GCS-BTZ)} black
hole ) has been studied in the context of the
higher-derivative/curvature gravities
\cite{Kalo:93,Cho:98,Garc:03,Krau:05b,Krau:05a,Solo:05a,Saho:06}.
And, it is found that the usual thermodynamic quantities of the BTZ
black hole, like as the black hole's mass, angular momentum, and
entropy are modified as
\begin{eqnarray}
M&=&m+x j/l, ~J=j+xlm, \label{MJ:general} \\
S&=&\frac{2 \pi r_+}{4 G \hbar}+x \frac{2 \pi r_-}{4 G \hbar},
\label{S:old}
\end{eqnarray} which shows some $mixings$ between the original BTZ
black hole's mass $m$ and angular momentum $j$, and also some
deviation, proportional to the $inner$-horizon's area, from the
usual Bekenstein-Hawking form \cite{Hawk:71,Beke:73} in the entropy
\cite{Garc:03,Krau:05a,Solo:05a,Saho:06}.
Here, the parameter $x$ is proportional to the gravitational
Chern-Simons coupling constant. These modifications would be the
results of the gravitational Chern-Simons dressing in the $AdS$
space, which have been absent in the usual topologically massive
gravity with $\La=0$.

But, that does not change much about the physical contents of the
usual BTZ black hole when the parameter $x$ is not large enough,
more exactly when it is smaller than a critical value of the
coupling constant. In fact, there is a good agreement in the entropy
(\ref{S:old}) with the statistical entropy, based on the conformal
field theory (CFT) for the Virasoro algebras at the spatially
infinite boundary \cite{Krau:05a,Solo:05a}, as in the usual BTZ
black hole systems \cite{Stro:98,Birm:98,Park:04a}.

However, for large values of the gravitational Chern-Simons coupling
where the modification terms dominate the original terms some exotic
behaviors occur, like as the roles of mass and angular momentum are
interchanged and its black hole entropy depends more on the
$inner$-horizon area than the outer one. Actually, similar phenomena
have been also known for some time in several other contexts
\cite{Carl:91}
where the masses and
angular momenta are {\it completely} interchanged and the black hole
entropies depend $only$ on the areas of the inner horizon ( I have
called these kinds of black holes as the {\it exotic black holes}
\cite{Park:06} ),
in completely contrast to the Bekenstein-Hawking entropy formula
\cite{Hawk:71,Beke:73}. This looks similar to the suggestion in Ref.
\cite{Lars:97}.

 But a basic $physical$ problem of those approaches is that the second law of
thermodynamics is not guaranteed with their entropy formulae,
in contrast to the Bekenstein-Hawking form \cite{Hawk:71}; actually,
without the guarantee of the second law, there is no justification
for identifying the entropies, even though they satisfy the first
law, with the inner-horizon areas \cite{Beke:73}. Moreover, those
entropies do $not$ agree with the statistical entropies, in contrast
to a good agreement for small values of the gravitational
Chern-Simons coupling, though this has not been well known in the
literatures.

In the usual system of black holes, the first law of thermodynamics
uniquely determines (up to an arbitrary constant) the black hole
entropy with a given Hawking temperature $T_{+}$ and chemical
potential for the outer (event) horizon $r_+$. In this context,
there is no choice in the entropy of the GCS-BTZ black hole,
other than (\ref{S:old}), which is problematic for large values of
$x$.

But recently, I have proposed a new entropy for
the case of
the exotic black holes \cite{Park:06}, which corresponds to the $|x|
\rightarrow \infty$ limit,
such as the entropy has the usual Bekenstein-Hawking form, which is
proportional to the area of the outer horizon.
And I have found that the new entropy formula have a good
agreement with the statistical entropy, based on the CFT at the
spatial infinity.

In this paper I
argue
that the new approach can be generalized to large but {\it
finite} values of $x$ also: By considering the characteristic
angular velocity and temperature as those of the inner horizon, a
new entropy formula is found from the first law of thermodynamics.
This new entropy agrees well with the statistical entropy. But, for
small values of $x$, the system behaves like as an ordinary BTZ
black hole with the characteristic angular velocity and temperature
as those of the outer horizon with the known entropy formula
(\ref{S:old}), which agrees with the statistical entropy as well.
So, I
argue that there are two different phases of the GCS-BTZ black hole,
depending on its gravitational Chern-Simons coupling constant. For
each phase, the second law
of the
thermodynamics is guaranteed and there are good agreements with the
statistical entropies.

The plan of this paper is as follows.

In Sec. II, I consider the thermodynamics of the GCS-BTZ black hole
and find the new entropy formula for the large gravitational
Chern-Simons coupling $|\hat{\beta}|>1$, as well as the usual
entropy formula for the small coupling $|\hat{\beta}|<1$, from a new
re-organization of the first law of thermodynamics.
I study also the
Smarr formula and find the same form as in the usual BTZ black holes
without the gravitational Chern-Simons term.

In Sec. III, the statistical entropy, based on the holographic
anomalies, is considered, and I find
 perfect agreements with the thermodynamic entropies that have been
 studied in Sec. II, for the $whole$ range of the gravitational Chern-Simons coupling.
 The new entropy formula, as well as the
 ordinary one, is
 supported by the CFT approach, which is
 robust in the context of the AdS/CFT correspondence.

In Sec. IV, the classical symmetry algebra approach, based on the
Chern-Simons formulation of three-dimensional gravity, is considered
for comparison with the holographic anomaly approach of Sec. III,
and I find $exact$ agreements between them. This provides a
non-trivial check of the AdS/CFT-correspondence, in the presence of
higher-derivative terms in the gravity action.
In order to ensure that the
exact {\it factor} matching even with the gravitational Chern-Simons
term is a solid result, by carefully fixing the subtleties involving
the normalization differences between the different bases and
conventions in the literatures, I include some details of the
computations and useful formulae in Appendix {\bf A}.

In Sec. V, I conclude with several discussions.

In Appendix {\bf B}, I briefly review on the derivation of the Cardy
formula and its higher-order corrections for completeness.

I shall omit the speed of light $c$ and the Boltzman's constant
$k_B$ in this paper for convenience, by adopting the units of
$c\equiv 1,~k_B \equiv 1$. But, I shall keep the Newton's constant
$G$ and the Planck's constant $\hbar$ in order to clearly
distinguish the quantum (gravity) effects with the classical ones.\\

\begin{section}
{Thermodynamics of the GCS-BTZ black hole}
\end{section}

\label{2}
\begin{subsection}
{The BTZ black hole in gravitational-Chern-Simons-corrected gravity}
\end{subsection}

The (2+1)-dimensional gravity with a gravitational Chern-Simons term
and a negative cosmological constant $\Lambda=-1/l^2$ is described
by the action on a manifold ${\cal M}$
\cite{Dese:82,Dese:91}
[ omitting some possible boundary terms ]
\begin{eqnarray}
\label{EHGCS} I_{g}=\frac{1}{16 \pi G} \int_{\cal M} d^3 x \sqrt{-g}
\left( R +\frac{2} {l^{2}}\right)+I_{GCS},
\end{eqnarray}
where the gravitational Chern-Simons term is given by \footnote{
Note that the $dimensioless$ coupling constant $\hat{\beta}=x$ is
related to the one used in Refs. \cite{Dese:82,Dese:91}
as
$\hat{\beta}=-1/( \mu l)$, in Ref. \cite{Solo:05a} as
$\hat{\beta}=-\beta_S/l$, and in Ref. \cite{Krau:05a} as
$\hat{\beta}=-32 \pi G \beta_{KL}/l$. } [ the Greek letters
($\mu,\nu,\alpha, \cdots$) denote the space-time indices and Latin
($a,b,c, \cdots$) denote the internal Lorentz indices; I take the
metric convention $\eta_{ab}$=diag$(-1,1,1)$ for the internal
Lorentz indices, and the indices are raised and lowered by the
metric $\eta_{ab}$ ]
\begin{eqnarray}
\label{GCS}
 I_{GCS}=\frac{\hat{\beta} l}{64 \pi G} \int_{\cal M} d^3 x
~\epsilon^{\mu \nu \alpha} \left( R_{ab \mu \nu}
{\omega^{ab}}_\alpha +\frac{2}{3}~ {\omega^b}_{c_\mu} {\omega^c}_{a
\nu} {\omega^a}_{b\alpha} \right).
\end{eqnarray}
Here, the spin-connection $1-form$ ${\omega^a}_b={\omega^a}_{b \mu}
dx^{\mu},~\omega_{ab\mu}=-\omega_{ba\mu}$ is determined by the
torsion-free condition
\begin{\eq}
\label{T=0:comp}
 d e^a +{\omega^a}_b \wedge e^b=0
\end{\eq}
with the dreibeins 1-form $e^a={e^a}_{\mu} dx^{\mu}$, and the
curvature is then $R_{ab \mu \nu}=\pa_{\mu}\omega_{ab
\nu}+{{\om_a}^c}_{\mu} \om_{cb\nu}-(\m \lra \n)$. [ I take the same
definitions as in Ref. \cite{Krau:05a} for the curvature 2-form
$R_{ab}=(1/2)R_{ab\m \n}~dx^{\m} \wedge dx^{\n}$ and the
spin-connection 1-form $\om_{ab}$. Some useful formulae are
summarized in Appendix {\bf A}. ] Note that $I_{GCS}$ is of the
third-derivative order, rather than the second as in the
Einstein-Hilbert term, so this is the first higher-derivative
correction in three-dimensional spacetimes.

The resulting equations of motion, by varying $I_{g}$ of
(\ref{EHGCS}) with respect to the metric\footnote{ The variations of
$I_{GCS}$ depends only the metric, though it does not look clear at
first sight, as $\de I_{GCS}=(l\hat{\be}/{8 \pi G}) \int_{\cal M}
d^3x \sqrt{-g}~ C^{\m \n} \de g_{\m \n}$ \cite{Dese:82,Krau:05a}. },
are
\begin{\eq}
\label{eom}
 R^{\mu \nu}-\f{1}{2} g^{\m \n} R -\f{1}{l^2} g^{\m \n}
=\hat{\be} l C^{\m \n},
\end{\eq}
where the Cotton tensor $C^{\m \n}$ is defined by
\begin{\eq}
C^{\mu \nu}= \f{1}{\sqrt{-g}} \epsilon^{\mu \rho \sigma}
\nabla_{\rho} ({R^{\nu}}_{\sigma}-\f{1}{4} {\delta^{\nu}}_{\sigma} R
),
\end{\eq}
which is traceless and covariantly conserved \cite{Dese:82}. From
the fact that the Einstein equation (\ref{eom}) gives a constant
curvature scalar $R=-6/l^2$, the equation (\ref{eom}) can be further
reduced to
\begin{\eq}
\label{eom2}
 R^{\mu \nu}
&=&\f{2}{l^2} g^{\m \n}+\hat{\be} l C^{\m \n} \no \\
&=&\f{2}{l^2} g^{\m \n}+\f{\hat{\be} l}{\sqrt{-g}} \epsilon^{\mu
\rho \sigma} \nabla_{\rho} {R^{\nu}}_{\sigma}.
\end{\eq}

It would be a non-trivial task to find the general black hole
solutions for the third-derivative-order equations\footnote{
Recently, a non-trivial two-parameter family of black hole solutions
have been found \cite{Mous:03}, but it does not seem that its
properties have been fully elucidated yet.}. However, there is a
trivial solution, e.g., the BTZ solution because it satisfies the
equation (\ref{eom2}) trivially with $C^{\m \n}= \epsilon^{\mu \rho
\sigma} \nabla_{\rho} {R^{\nu}}_{\sigma}/{\sqrt{-g}}=0$
\cite{Kalo:93}. This looks like a too-trivial situation which does
not have any higher-derivative effect of the gravitational
Chern-Simons term. But actually this is not the case, as we will
see, since there are some non-trivial shifts in the physical
parameters of the black hole \cite{Garc:03,Krau:05a,Solo:05a};
actually, the BTZ solution is rich enough to show some genuine
effect of the gravitational Chern-Simons term. So, I concentrate
hereafter only the BTZ solution, which is given by the metric
\cite{Bana:92}
\begin{eqnarray}
\label{BTZ}
 ds^2=-N^2 dt^2 +N^{-2} dr^2 +r^2 (d \phi +N^{\phi}
dt)^2
\end{eqnarray}
with
\begin{eqnarray}
\label{BTZ:N}
 N^2=\frac{(r^2-r_+^2) (r^2-r_-^2)}{l^2 r^2},~~
N^{\phi}=-\frac{r_+ r_-}{l r^2}.
\end{eqnarray}
Here, $r_+$ and $r_-$ denote the outer and inner horizons,
respectively.

In the absence of the gravitational Chern-Simons term, the conserved
mass and angular momentum of the black hole are given by
\begin{eqnarray}
\label{mj}
 m=\frac{r_+^2 +r_-^2}{8 Gl^2},~~j=\frac{2 r_+ r_-}{8Gl },
\end{eqnarray}
respectively. Note that these parameters satisfy the usual
mass/angular momentum inequality $m^2 \geq j^2/l^2$, in order that
the horizon exists or the conical singularity is not naked and the
equality holds for the extremal black hole, where the inner and
outer horizons overlap.

But, in the presence of the gravitational Chern-Simons term, it has
been found that these ``bare'' parameters are shifted as
(\ref{MJ:general})\footnote{ This has been computed in several
different approaches, e.g., the super-angular momentum's in Ref.
\cite{Mous:03}, the quasi-local method's in Ref. \cite{Garc:03}, the
ADM's in Ref. \cite{Dese:05},
the holography's in Refs.
\cite{Krau:05a,Solo:05a}. But, they all give the same result. },
i.e.,
\begin{eqnarray}
\label{M_J}
 M=m+\hat{\be} j/l, ~~ J=j+ \hat{\be}l m \label{MJ},
\end{eqnarray}
respectively and these modifications would be the results of
gravitational Chern-Simons term in $AdS$ space. One remarkable
result of these modifications is that the usual mass/angualr
momentum inequality is not valid generally
\begin{eqnarray}
\label{M_bound}
 M^2 -J^2/l^2 =(1-\hat{\be}^2) (m^2-j^2/l^2),
\end{eqnarray}
but it depends on  the values of the gravitational Chern-Simons
coupling constant $\hat{\be}$: For small values of coupling
$|\hb|<1$, the usual inequality is preserved, i.e., $M^2  \geq
J^2/l^2$; however, for the large values of coupling $|\hb|>1$, one
has an {\it anomalous} inequality with an exchanged role of the mass
and angular momentum as $J^2/l^2 \geq M^2$; on the other hand, at
the critical value $|\hb|=1$, the modified mass and angular momentum
are ``always'' saturated, i.e., $M^2 = J^2/l^2$, regardless of
inequality of the bare parameters $m$
and $j$.\\

\begin{subsection}
{Black hole thermodynamics}
\end{subsection}

Since the solution (\ref{BTZ}) has the same form of the metric as
the usual BTZ solution, it has the same form of the Hawking
temperature and angular velocity of the outer (event) horizon $r_+$
as the BTZ also
\begin{eqnarray}
T_+=\left. \frac{\hbar \kappa}{2 \pi} \right|_{r_+}=\frac{\hbar
(r_+^2 -r_-^2)}{2 \pi l^2 r_+},~~\Omega_+=\left.-N^{\phi}
\right|_{r_+}=\frac{r_-}{l r_+}
\end{eqnarray}
with the surface gravity function $\kappa=\partial N^2/( 2 \partial
r)$.

Now, by considering the first law of thermodynamics as
\begin{eqnarray}
\label{first:old}
 \delta M=\Omega_+\delta J + T_+ \delta S
\end{eqnarray}
with $T_+$ and $\Omega_+$ as the characteristic temperature and
angular velocity of the system, one can easily determine the black
hole entropy as
\begin{eqnarray}
\label{S:old2}
 S=\frac{2 \pi r_+}{4 G \hbar}+\hb \frac{2 \pi r_-}{4
G \hbar}.
\end{eqnarray}
There is no other choice in the entropy in this usual context
\cite{Carl:91,
Solo:05a,Saho:06}. In fact,
this has been computed also in rather formal contexts, like as the
Euclidean method of conical singularity \cite{Solo:05a} and Wald's
formalism \cite{Saho:06}, but the same entropy has been obtained.

However, an inherent problem of all those approaches is that there
is {\it no} general proof about the second law of thermodynamics
when higher-derivative/curvature terms are included in general
\cite{Jaco:94}.
In our case of (\ref{S:old2}),
there are two contributions: One is the usual Bekenstein-Hawking
term
\begin{\eq}
\label{S:BH}
 S_{BH}=\frac{2 \pi r_+}{4 G \hbar},
\end{\eq}
 which guarantees the second law
from Hawking's area theorem \cite{Hawk:71,Beke:73}, which saying the
increase of the area of the outer horizon ${\cal A}_+=2 \pi r_+$.
Another term is proportional to the inner-horizon area ${\cal A}_-=2
\pi r_-$ and this comes from the gravitational Chern-Simons term.
But, in this second part, the second law would be questionable since
some of the basic assumptions for the Hawking's area theorem, i.e.,
cosmic censorship conjecture might not be valid for the inner
horizon, in general. Moreover, the usual instability of the inner
horizon makes it difficult to apply the Raychaudhuri's equation to
get the area theorem, even without worrying about other assumptions
for the theorem; actually, this seems to be the situation that
really occurs in our GCS-BTZ black holes also
\cite{Stei:94,Bala:04}.

But, there is a novel situation where the total entropy
(\ref{S:old2}) still satisfies the second law, though all its
constituents do not. This is the case where the usual
Bekenstein-Hawking term dominates the exotic term proportional to
${\cal A}_{-}$: Since $r_+\geq r_-$ is always satisfied, this
condition is equivalent to $|\hb|<1$. Actually, this is the case
where the usual mass/angular momentum inequality holds, as I have
shown in the previous sub-section II.A, the system behaves as an
ordinary BTZ black hole, though there are some shifts in the mass,
angular momentum, and entropy.

On the other hand, for large values of coupling $|\hb|>1$, where the
exotic term dominates the Bekenstein-Hawking term, the above
argument does not guarantee the second law of thermodynamics
generally. Then, without the guarantee of the second law of
thermodynamics, there is no justification for identifying entropy
(\ref{S:old2}), even though it satisfies the first law of
thermodynamics (\ref{first:old}), and its characteristic temperature
and angular velocity have the usual identifications \cite{Beke:73}.

So, in order to avoid the problem for the large couplings, we need
another form of the entropy which is dominated by a term $linearly$
proportional to the outer horizon area ${\cal A}_+$, following the
Bekenstein's general argument \cite{Beke:73}, which should be valid
in my case also.
Recently, I have studied the extreme limit $|\hb| \ra \infty$ of the
system and found that the new entropy formula can be determined by a
new re-organization of the first law of thermodynamics; here, I
consider its
generalization to my case. A crucial fact for the new formulation is
by observing the following identities in the BTZ system
\begin{\eq}
\label{id1}
 \de m&=&\Om_+ \de j +\f{(r_+^2 -r_-^2)}{2 \pi l^2 r_+}
\left(\f{2 \pi \de
r_+}{4 G}  \right)\\
\label{id2}
 &=&\Om_- \de j +\f{(r_-^2 -r_+^2)}{2 \pi l^2 r_-} \left(\f{2
\pi \de r_-}{4 G} \right),
\end{\eq}
where
\begin{\eq}
\label{Om-} \Omega_-=\left.-N^{\phi} \right|_{r_-}=\frac{r_+}{l r_-}
\end{\eq}
is the angular velocity for the inner horizon; these identities show
a symmetry between $r_+$ and $r_-$, which would reflect the symmetry
in the metric (\ref{BTZ}), (\ref{BTZ:N}) and the bare parameters
(\ref{mj}).

Then, the first identity (\ref{id1}) produces the usual first law of
thermodynamics with the Hawking temperature $T_+$, angular velocity
$\Om_+$ for the outer horizon, and Bekenstein-Hawking entropy
$S_{BH}$:
\begin{\eq}
\de m=\Om_+ \de j +T_+ \de S_{BH}.
\end{\eq}

The second identity (\ref{id2}) is an interesting re-arrangement of
the first identity by replacing $r_+$ with $r_-$; this would be
remarkable since the first law does $not$ uniquely determine (up to
a constant) the black hole entropy, as well as the characteristic
temperature and angular velocity, in contrast to usual belief;
actually, the second identity (\ref{id2}) implies that the system
can be also considered as a black hole with the entropy
\begin{\eq}
S_-=\f{2 \pi r_-}{4 G \hbar},
\end{\eq}
which is proportional to the inner-horizon area ${\cal A}_{-}$, and
the characteristic temperature\footnote{ I have used the definition
of $\kappa$ as $\nabla ^{\nu} (\chi ^{\mu} \chi_{\mu} )=-\kappa
\chi^{\nu}$ for the horizon Killing vector $\chi^{\mu}$ in order to
determine its {\it sign}, as well as its magnitude. }
\begin{eqnarray}
T_-=\left. \frac{\hbar \kappa}{2 \pi} \right|_{r_-}=\frac{\hbar
(r_-^2 -r_+^2)}{2 \pi l^2 r_-}
\end{eqnarray}
and angular velocity $\Om_-$ for the inner horizon:
\begin{\eq}
\label{2.22}
 \de m=\Om_- \de j +T_- \de S_{-}.
\end{\eq}
Here, the physical relevances of the parameters $T_-$ and $\Om_-$
are not clear. But, here and below, I use $T_-$, $\Om_-$ just for
convenience in identifying the new entropy, from the ``assumed''
first law of thermodynamics (\ref{2.22}).\footnote{ The
positive-valued surface gravity and temperature with $T=\left|
\kappa_{-} /(2 \pi) \right| $ (as in Ref. \cite{Bala:04}) produces
an incorrect sign in front of the $TdS$ term in (\ref{first:new2}).
}

Now, let me consider, from (\ref{M_J}),
\begin{\eq}
\label{first:new1}
 \de M- \Om_{-} \de J =\de m- \Om_{-} \de j +
\hb( \de j/l-\Om_{-} l \de m),
\end{\eq}
instead of $\de M- \Om_{+} \de J$ in (\ref{first:old}). Then, it is
easy to see that the first two terms in the right hand side become
$T_{-} \de S_{-}$, by using the second identity (\ref{id2}) or  from
(\ref{2.22}). And also, the final two terms in the bracket become
$T_{-} \de S_{BH}$, by using the first identity (\ref{id1}) and
another identity
\begin{\eq}
\label{Om-+}
\Om_{-}=\Om_{+}^{-1} l^{-2}.
\end{\eq}
So, finally I find that (\ref{first:new1}) becomes a new
re-arrangement of the first law as
\begin{eqnarray}
\label{first:new2}
 \delta M=\Omega_-\delta J + T_- \delta S_{\new},
\end{eqnarray}
with the new black hole entropy
\begin{eqnarray}
S_{\new}=\frac{2 \pi r_-}{4 G \hbar}+\hb \frac{2 \pi r_+}{4 G
\hbar}. \label{S:new}
\end{eqnarray}

With the above new entropy formula, it is easy to see that the
previous argument for the second law of thermodynamics of
(\ref{S:old2}) in the small values of coupling $|\hb| <1$ can now be
applied to that of (\ref{S:new}) in the large values of coupling
$\hb > 1$.

On the other hand, for the large but ``negative'' values of coupling
$\hb <-1$, the entropy formula (\ref{S:new}) would $not$ guarantee
the second law of thermodynamics $nor$ the $positiveness$ of the
entropy: The entropy would ``decrease'' indefinitely, with the
negative values, as the outer horizon $r_+$ be increased, following
the area theorem. But, there is a simple way of resolution from the
new form of the first law (\ref{first:new2}). It is to
consider
\begin{eqnarray}
{S_{\new}} '&\equiv& -S_{\new}=-\frac{2 \pi r_-}{4 G \hbar}-\hb
\frac{2
\pi r_+}{4 G \hbar}, \label{S:new'} \\
{T_-}' &\equiv& -T_-=\frac{\hbar (r_+^2 -r_-^2)}{2 \pi l^2 r_-},
\end{eqnarray}
instead of $S_{\new},T_-$, and actually this seems to be the
$unique$ choice: One might consider ${S_{\new}} ''\equiv \frac{2 \pi
r_-}{4 G \hbar}-\hb \frac{2 \pi r_+}{4 G \hbar}$, but then the first
law (\ref{first:new2}) is $not$ satisfied.
\\

\begin{subsection}
{Smarr formula and its universality}
\end{subsection}

So far, I have
argued that there are two different phases of the GCS-BTZ black
hole, depending on its gravitational Chern-Simons coupling. The
physics is quite different in the two phases, having different
thermodynamic functions, $T_+, \Om_+,S$ for $|\hb|<1$ and
$T_-,\Om_-,S_{\new}$ for $|\hb|>1$. But, for each phase, the second
law, as well as the (assumed) first law of thermodynamics, is
guaranteed.

On the other hand, it is known that the bare BTZ black hole
satisfies the three-dimensional Smarr formula \cite{Cai:97}
\begin{\eq}
\label{Smarr+}
 m=\f{1}{2} T_+ S_{BH} +\Om_{+} j.
\end{\eq}
So, an interesting question would be whether this formula is
deformed in the presence of the higher-derivative/curvature terms in
the action; also, the study of this relation would be important in
that it could show some universal characteristics of the system, in
connection with other thermodynamic systems which look completely
different.

This would be a non-trivial question in the general
asymptotically-AdS space \cite{Gibb:05}.
But, in our
GCS-BTZ case, the Smarr formula (\ref{Smarr+}) is unchanged from
some magic of the system. The magic comes, first, from the following
identity, in addition to (\ref{Smarr+}),
\begin{\eq}
\label{Smarr-}
 m=\f{1}{2} T_- S_{-} +\Om_{-} j
\end{\eq}
and this can be considered as another re-arrangement of the
three-dimensional Smarr formula (\ref{Smarr+}), which has never been
considered in the literatures.
And also, by considering
(\ref{Om-+}) and $T_{-} \Om_{-}^{-1}=-T_+/l$, one has the identities
\begin{\eq}
\label{Smarr-:j}
 {j}/{l}&=&\f{1}{2} T_- S_{+} +l \Om_{-} m \\
\label{Smarr+:j}
    &=&\f{1}{2} T_+ S_{-} +l \Om_{+} m.
\end{\eq}
The first and second identities come from (\ref{Smarr+}) and
(\ref{Smarr-}), respectively.

Then, from all these magical identities, one can easily find the
following Smarr formulae for the
gravitational-Chern-Simons-corrected mass and angular momentum, M
and J, respectively
\begin{\eq}
\label{Smarr+:GCS}
 M&=&\f{1}{2} T_+ S_{} +\Om_{+} J, \\
 \label{Smarr-:GCS}
 M&=&\f{1}{2} T_- S_{\new} +\Om_{-} J \\
 &=&\f{1}{2} {T_-}' {S_{\new}}' +\Om_{-} J,
\end{\eq}
by considering  (\ref{Smarr+}) and (\ref{Smarr+:j}), and
(\ref{Smarr-}) and (\ref{Smarr-:j}), respectively. Here,
(\ref{Smarr+}) and (\ref{Smarr+:GCS}) describe the black holes with
$|\hb|<1$ since $T_+$ and $\Om_+$ are considered as the
characteristic parameters of the system. Similarly, (\ref{Smarr-})
and (\ref{Smarr-:GCS}) describe those with $|\hb|>1$.

So, I have found that the two Smarr formulae (\ref{Smarr+}) and
(\ref{Smarr-}) extend to the GCS-BTZ black hole with the corrected
$M,~J$, and the entropies $S$, $S_{\new}$, or
${S_{\new}}'$.
However, it is not clear whether the $covariance$ of Smarr formula
is just a result of the speciality of
the gravitational Chern-Simons term or there are other deep reasons.\\

\begin{section}
{Statistical entropy: The holographic anomaly approach}
\end{section}

In the usual context of the AdS/CFT correspondence \cite{Ahar:00},
the central charges for the CFT on the asymptotic AdS boundary are
identified by evaluating the anomalies of the CFT effective action,
from the regularized bulk gravity action
\cite{Henn:98,Hyun:99,Bala:99}.

Recently, the approach has been applied to the action (\ref{EHGCS}),
and it is found that there are anomalies in the expectation values
of the boundary stress tensor $T_{ij}={2} {\de I_g
[\ga^{ij}]}/{\sqrt{-\ga}}{\de \ga^{ij}}$, for the boundary metric
$ds^2 =\ga_{ij} dx^i dx^j \simeq -r^2 dx^{+} dx^{-}$ with $r$ taken
to infinity,
\begin{\eq}
\label{anomaly} \left< T_{++} (x^+)\right>=-\f{\hbar \hat{c}^+}{24
\pi},~~\left< T_{--} (x^+)\right>=-\f{\hbar \hat{c}^-}{24 \pi},
\end{\eq}
with the central charges [ I follow the conventions of
\cite{Bala:99} ]
\begin{\eq}
\hat{c}^{\pm}=\gamma^{\pm} \frac{3 l}{2G \hbar}
\label{c:anomaly}
\end{\eq}
with $\gamma^{\pm}=1\pm {\hb}$ for the right/left-moving sectors
with the superscripts $+$ and $-$, respectively. Here, I have
defined $\hat{c}^{\pm} \sim O(\hbar^{-1})$ as the quantum-mechanical
central charges of the boundary CFT, due to (quantum) anomaly
$\left< {2} \ga_{ij} {\de \log Z}/{\sqrt{-\ga}}{\de \ga^{ij}}
\right>=-\hat{c} R^{(2)}/(24 \pi)\sim O(\hbar^{-1})$ such as they
contain the Planck's constant $\hbar$, intrinsically. However, the
bulk computation of the holographic anomaly has no $\hbar$ because
it just uses the classical action and equations of motion,
e.g., $ T_{\pm \pm} =- {c}^{\pm}/(24 \pi)\sim O(1)$, with classical
numbers $c^{\pm} \sim O(1)$.
Note that the quantum-mechanical central charges $\hat{c}^{\pm}$
defined in this way have the correct ``$1/\hbar$''-factor for
the semiclassical black hole entropy, like as the Bekenstein-Hawking
entropy (\ref{S:BH}), via the Cardy formula \ci{Card:86,Park:02}. It
seems that this $1/\hbar$-factor is closely related to that of the
black hole entropy in Euclidean action approach \ci{Brad:90}, where
$1/\hbar$-factor occurs from the zero-loop approximation of the
effective action $\log Z \approx -I_{\eff} \hbar^{-1}$.

By considering (\ref{anomaly}) as the anomalous transformations of
the boundary stress tensors under the diffeomorphism $\de
x^{\pm}=-\xi^{\pm}(x^{\pm})$,
\begin{\eq}
\de_{\xi^{+}}T_{++}&=&2 \pa_+ \xi^+ T_{++} +\xi^+ \pa_+
T_{++}-\f{\hbar \hat{c}^+}{24 \pi} \pa^3_+ \xi^+ \no \\
&=&\f{1}{i}[T_{++}, \hat{L}^+[\xi^+]], \no \\
 \de_{\xi^{-}}T_{--}&=&2 \pa_- \xi^-
T_{--} +\xi^- \pa_- T_{--}-\f{\hbar \hat{c}^-}{24 \pi} \pa^3_- \xi^-
\no \\
&=&\f{1}{i}[T_{--}, \hat{L}^-[\xi^-]]
\end{\eq}
with the generators
\begin{\eq}
\hat{L}^{\pm}[\xi^{\pm}]=\f{1}{\hbar} \oint dx^{\pm} T_{\pm \pm}
\xi^{\pm} (x^{\pm})+\f{\hat{c}^{\pm}}{24 },
\end{\eq}
one can obtain a pair of quantum Virasoro algebras
\begin{\eq}
[\hat{L}^{\pm}_{m},\hat{L}^{\pm}_{n}]=(m-n) \hat{L}^{\pm}_{m+n}+\f{
\hat{c}^{\pm}}{12 } m (m^2-1) \de_{m+n,0}
\end{\eq}
for a monochromatic basis $\xi^{\pm}=e^{imx^{\pm}}$ with the integer
numbers $m$, $n$. Here I note that this reduces to the usual result
for the holographic {\it conformal} anomaly in the $\hb \ra 0$ limit
\cite{Henn:98,Hyun:99,Bala:99}, whereas $\hb$-dependent terms come
from the holographic {\it gravitational} anomaly due to the
gravitational Chern-Simons term \cite{Krau:05a,Solo:05a}.

Now, let me consider the ground state Virasoro generators, expressed
in terms of the black hole's mass and angular momentum:
\begin{\eq}
\label{ground}
 \hat{L}^{\pm}_{0}&=&\frac{lM \pm J}{2\hbar} +\f{
\hat{c}^{\pm}}{24
}\no \\
&=&\gamma^{\pm} \frac{(lm \pm j)}{2\hbar} +\f{ \hat{c}^{\pm}}{24 }.
\end{\eq}

With the Virasoro algebras of $\hat{L}^{\pm}_{m}$ in the standard
form, which are defined on the {\it plane}, one can use the Cardy
formula for the asymptotic states
\cite{Card:86,Carl:99,Park:02,Kang:04,Park:04b}
\begin{eqnarray}
\label{Cardy}
 \mbox{log}~ \rho (\hat{\Delta}^{\pm}) \simeq 2 \pi
\sqrt{ \f{1}{6} \left(\hat{c}^{\pm}-24
  \hat{\Delta}^{\pm}_{\hbox{\scriptsize
  min}}\right)\left(\hat{\Delta}^{\pm}-\frac{\hat{c}^{\pm}}{24}\right) },
\end{eqnarray}
where $\hat{\Delta}^{\pm}$ are the eigenvalues, called conformal
weights, of the operator $\hat{L}_0$ for black-hole quantum states
$|\hat{\Delta}^{\pm}\rangle$, and
$\hat{\Delta}^{\pm}_{\hbox{\scriptsize min}}$ are their minimum
values. Here, I note that the above Cardy formula, which comes from
the saddle-point approximation of the CFT partition function on a
torus, is valid only if the following two conditions are satisfied:
\begin{\eq}
\label{cond1}
\f{24 {\hat \De}^{\pm}_{\eff}}{\hat{c}^{\pm}_{\eff}} &\gg& 1, \\
\label{cond2}
 \hat{c}^{\pm}_{\eff} {\hat \De}^{\pm}_{\eff} &\gg& 1,
\end{\eq}
where $\hat{\Delta}^{\pm}_{\hbox{\scriptsize
eff}}=\hat{\Delta}^{\pm} -\hat{c}^{\pm}/{24}$ and
$\hat{c}^{\pm}_{\hbox{\scriptsize
eff}}=\hat{c}^{\pm}-24\hat{\Delta}^{\pm}_{\hbox{\scriptsize min}}$
are the effective conformal weights and central charges,
respectively; from the first condition, the higher-order correction
terms are exponentially suppressed as $e^{- 2 \pi \ep^{\pm}
(\hat{\Delta}^{\pm}-\hat{\Delta}^{\pm}_{\hbox{\scriptsize min}})}$
with $\ep^{\pm} \equiv 24 {\hat
\De}^{\pm}_{\eff}/\hat{c}^{\pm}_{\eff}$; from the second condition,
the usual saddle-point approximation is reliable, i.e.,
$\rho(\hat{\Delta}^{\pm})$ dominates in the partition function (see
Appendix {\bf B} for the details).

Then, the statistical entropy for the asymptotic states becomes
\begin{\eq}
S_{\stat}&=&\mbox{log}~ \rho (\hat{\Delta}^{+})+\mbox{log}~ \rho
(\hat{\Delta}^{-}) \no \\
&=&\f{\pi}{4 G \hbar} |\ga^+ (r_+ + r_-)|+\f{\pi}{4 G \hbar} |\ga^-
(r_+ - r_-)| \no \\
&=&\f{\pi}{4 G \hbar} (|\ga^+| +|\ga^-|) r_+ +\f{\pi}{4 G \hbar}
(|\ga^+|-|\ga^-|)r_-~, \label{S:stat}
\end{\eq}
where I have chosen $\hat{\Delta}^{\pm}_{(\hbox{\scriptsize
min})}=0$, as usual \cite{Stro:98,Birm:98,Park:04a}; from
(\ref{ground}), this corresponds to the $AdS_3$ vacuum solution with
$m=-1/(8G)$ and $ j=0$, in the usual context, but it has a permanent
rotation as well, in the new context \cite{Krau:05a},
\begin{\eq}
M=-\f{1}{8G},~~J=-\f{l \hb}{8G}.
\end{\eq}
Note that the correct ``$1/\hbar$''-factor for the semiclassical
black hole entropy comes from the appropriate recovering of $\hbar$
in (\ref{c:anomaly}) and (\ref{ground}).  According to the
conditions of validity (\ref{cond1}), (\ref{cond2}), this entropy
formula is valid only when both of the two conditions
\begin{\eq}
\label{cond3}
&&(r_+\pm r_-) \gg l, \\
\label{cond4}
 &&(r_+ \pm r_-) \gg \hbar G
\end{\eq}
are satisfied. The usual semiclassical limit of large black hole
(area), in which the back-reaction of the emitted radiation from the
black hole is neglected \cite{Pres:91}
and so the
thermodynamic entropy formula (\ref{S:old2}) and (\ref{S:new}) from
the first law can be reliable, agrees with the condition
(\ref{cond4}). So, there would be no obstacles to compare the
statistical entropy (\ref{S:stat}) with the thermodynamical one.
Note that from another condition (\ref{cond3}) we are considering a
more restricted class of black hole systems\footnote{ At this state,
the condition of large central charges $\hat{c}^{\pm} \gg 1$, i.e.,
$l \gg \hbar G$ \cite{Stro:98}, which would be related to the
leading supergravity approximation of AdS/CFT correspondence
\cite{Ahar:00}, is not needed yet.}, though this does not seem to be
needed, in general.

Now, let me consider the following four cases, depending on the
values of $\hb$: (a). $|\hb| < 1$, (b). $\hb > 1$, (c). $\hb <-1$,
and
(d). $|\hb| = 1$.\\

(a). $|\hb| < 1$: In this case, I have $|\ga^{\pm}|=\ga^{\pm}$, and
the statistical entropy (\ref{S:stat}) becomes
\begin{eqnarray}
\label{S_stat:old2}
 S_{\stat}=\frac{2 \pi r_+}{4 G \hbar}+\hb \frac{2 \pi r_-}{4
G \hbar}
\end{eqnarray}
from $\ga^++\ga^-=2,~\ga^+-\ga^-=2 \hb$. This agrees exactly with
the usual entropy formula (\ref{S:old2}). And, this is the case
where $\hat{c}^{\pm}$ and $\hat{\Delta}^{\pm} -\hat{c}^{\pm}/{24}$
are positive definite such as the Cardy formula (\ref{Cardy}) has a
well-defined meaning. In the gravity side also, it shows the usual
behavior with the ``positive'' mass and angular momentum, satisfying
the normal inequality $M^2 \ge J^2/l^2$.\\

(b). $\hb > 1$: In this case, I have
$|\ga^{+}|=\ga^{+},~|\ga^{-}|=-\ga^{-}$, and so the statistical
entropy (\ref{S:stat})
becomes
\begin{eqnarray}
\label{S_stat:new}
 S_{\stat}=\frac{2 \pi r_-}{4 G \hbar}+\hb \frac{2 \pi r_+}{4
G \hbar}.
\end{eqnarray}
This agrees exactly with the new entropy formula (\ref{S:new}),
which guarantees the second law of thermodynamics even in this case.
And, this is the case where there is some abnormal change of the
role of the mass and angular momentum due to $J^2/l^2\ge M^2 $, even
though $M$ and $J$ both are positive definite, as usual. Moreover,
in the CFT side also, this is not the usual system because
$\hat{c}^{-}=\ga^- 3 l/(2G \hbar)$ and $\hat{\Delta}^{-}
-\hat{c}^{-}/{24}=\ga^-(ml -j)/2\hbar$ are $negative$ valued, though
their self-compensations of the negative signs produce the $real$
and $positive$ statistical entropy. The application of the Cardy
formula to the case of negative $\hat{c}^{-}$ and $\hat{\Delta}^{-}
-\hat{c}^{-}/{24}$ might be questioned, due to the existence of
negatives-norm states with the usual condition $\left.
\hat{L}^-_n|\hat{\Delta}^{-}\right>=0~(n>0)$ for the highest-weight
state $\left.|\hat{\Delta}^{-}\right>$. However, this problem can be
easily cured by considering another representation of the Virasoro
algebra with $\tilde{L}_{n}^-\equiv
-\hat{L}_{-n}^-,~\tilde{c}^-\equiv-\hat{c}^-$, and
$\tilde{L}_n^-|\tilde{\De}^- \left.\right>=0~(n>0)$ for the new
highest-weight state $|\tilde{\De}^-\left.\right>$ \cite{Bana:99b};
this implies that the Hilbert space need to be ``twisted'' in which
the whole states vectors be constructed from the twisted
highest-weight state $\left.|\hat{\Delta}^{+}\right>
\otimes\left.|\tilde{\De}^-\right>$. The formula (\ref{S:stat}),
which is invariant under this substitution, should be understood in
this context. On the other hand,
 if I take the limit $\hb \ra \infty$ , in which there is only the
 gravitational Chern-Simons term, this becomes the ``exotic'' black hole system that occur
 in several different contexts \cite{Carl:91,
Park:06}; however, note that this can $not$ be obtained from (\ref{S_stat:old2}). \\

(c). $\hb <-1$: In this case, I have
$|\ga^{+}|=-\ga^{+},~|\ga^{-}|=\ga^{-}$, and the statistical entropy
(\ref{S:stat}) becomes
\begin{eqnarray}
\label{S_stat:new2}
 S_{\stat}=-\frac{2 \pi r_-}{4 G \hbar}-\hb \frac{2 \pi r_+}{4
G \hbar}.
\end{eqnarray}
Note that this is positive definite, and this should be the case
from its definition
$S_{\stat}=\log(\rho(\hat{\De}^+_0)\rho(\hat{\De}^-_0))\ge 0$ for
the number of possible states $\rho (\hat{\De}^{\pm}_0) \geq 1$.
This agrees exactly with the new entropy formula (\ref{S:new'}),
which guarantees the second law.
And, this is the case where $M$ $can$ be negative and $J$ has the
opposite direction to the bare one $j$, in contrast to the positive
definite $M$ and $J$ in the cases of (a) and (b), as well as the
anomalous inequality $J^2/l^2 \geq M^2$. In the CFT side,
$\hat{c}^{+}$ and $\hat{\Delta}^{+} -\hat{c}^{+}/{24}$ become
negative-valued, now, and I need to twist this right-moving sector,
rather than the left-moving one as in the case of (b),
$\tilde{L}_{n}^+\equiv
-\hat{L}_{-n}^+,~\tilde{c}^+\equiv-\hat{c}^+$, and
$\tilde{L}_n^+|\tilde{\De}^+ \left.\right>=0~(n>0)$ for the twisted
highest-weight state $\left.|\tilde{\Delta}^{+}\right>
\otimes\left.|\hat{\De}^-\right>$.
\\

(d). $|\hb| = 1$: In this case, one of $\ga^{\pm}$ vanishes, i.e.,
$\ga^+=0,~\ga^-=2$ for $\hb=-1$, and $\ga^+=2,~\ga^-=0$ for $\hb=1$.
The statistical entropy becomes
\begin{eqnarray}
\label{S_stat:new3}
 S_{\stat}=\frac{2 \pi}{4 G \hbar}(r_+ - r_-)~~~~(\hb=-1),\\
 \label{S_stat:new4}
 S_{\stat}=\frac{2 \pi}{4 G \hbar}(r_+ + r_-)~~~~(\hb=+1).
\end{eqnarray}
Note that (\ref{S_stat:new4}) can be reproduced from
(\ref{S_stat:old2}) and (\ref{S_stat:new}), but (\ref{S_stat:new3})
from (\ref{S_stat:old2}) and (\ref{S_stat:new2}). So, statistical
entropies (\ref{S_stat:new3}) and (\ref{S_stat:new4}) agree exactly
with the usual entropy formula (\ref{S_stat:old2})$\sim$
(\ref{S_stat:new2}). As I have remarked previously in Sec. II. (b),
this is the case where the mass/angular momentum inequality
saturates $M^2=J^2/l^2$, regardless of $m$ and $j$. In fact, they
satisfy
\begin{\eq}
M=\pm J/l=\f{(r_+ \pm r_-)^2}{8 Gl^2} \ge 0
\end{\eq}
for $\hb=\pm 1$, respectively. So, for non-extremal bare black holes
with $r_+ >r_-$, the mass $M$ is positive definite, but $J$ changes
its direction for $\hb=-1$. For extremal bare black holes with
$r_+=r_-$, one has $M=J=0$, as well as $S_{\stat}=S=0$ satisfying
the Nernst formulation of the third law of thermodynamics
\cite{Hawk:94}
for $\hb=-1$, whence $M=J=(G
\hbar/(2 \pi^2 l^2)) S_{\stat}=r_+^2/(2 G l^2)>0$ without satisfying
the third law for $\hb=1$, as in all other cases of (a)$\sim$(c) and
in the usual Kerr black hole \cite{Wald:97}. But, there are some
subtleties about this in the fully corrected entropies; see Sec. V
about this issue.

In summary, I have found exact agreements between the
thermodynamical black hole entropies which have been evaluated in
the bulk (AdS) gravity side and the CFT entropies in the asymptotic
boundary, for the $whole$ range of the coupling constant $\hb$.
So, the new entropy formula for the strong coupling $|\hb| >1$
seems to be supported by the CFT approach also. This reveals the
AdS/CFT correspondence in the sub-leading order with the
higher-derivative term of gravitational Chern-Simons, as well as in
the leading order with
the Einstein-Hilbert action.\\

\begin{section}
{Comparison with a classical symmetry algebra approach:
 Exact agreements}
\end{section}

 There is an alternative approach to compute the Virasoro
algebras with central charges. This is based on the {\it classical}
symmetry algebras of the asymptotic isometry of $AdS_3$
\cite{Brow:86,Bana:99,Oh:98,Park:98,Park:99}
\begin{\eq}
\label{Virasoro:class}
 \{{L}^{\pm}_{m},{L}^{\pm}_{n}\}^{*}\approx
i(m-n) {L}^{\pm}_{m+n}+\f{ i c^{\pm}}{12 } m (m^2-1) \de_{m+n,0},
\end{\eq}
with ``classical'' central charges $c^{\pm}$ and the Dirac bracket
$\{~,~\}^*$ \cite{Dira:64}.

It is well known that there is an exact agreement with the anomaly
based approaches of Sec. III by the mapping \cite{Park:02}, with the
appropriate recovering of $\hbar$,
\begin{\eq}
\label{c-c}
\hat{c}^{\pm}=\f{c^{\pm}}{\hbar},~~\hat{L}^{\pm}_m=\f{L^{\pm}_m}{\hbar},
\end{\eq}
in the absence of the gravitational Chern-Simons term
\cite{Henn:98,Hyun:99,Bala:99} \footnote{ The classical algebra with
the higher curvature terms was computed in Ref. \cite{Said:00} by
transforming the gravity action with the higher curvature terms into
the usual Einstein-Hilbert action with some auxiliary tensor matter
fields. The same central charges and Virasoro generators have been
obtained in the anomaly approach also recently \cite{Krau:05b}. But
the validity of Ref. \cite{Said:00} is unclear since there would be
non-trivial contributions in the generators $L_m^{\pm}$ and central
charges from the matter fields {\it in general}
\cite{Park:04a,Nats:00},
though the
agreement seems to be plausible in the context of AdS/CFT
\cite{Noji:99,Krau:05b}.}. So, the statistical entropy agrees with
the Bekenstein-Hawking entropy also. But, this is a quite
non-trivial fact, and actually this provides an $explicit$ check of
the AdS/CFT correspondence by comparing the classical data
$(c^{\pm},L^{\pm})$, which can be $directly$ computed, with the
quantum data $(\hat{c}^{\pm},\hat{L}^{\pm})$ in the anomaly
approach, which can be identified {\it only indirectly} through the
(conjectured) AdS/CFT-correspondence.

So, it would be interesting to consider the classical approach in
the presence of the gravitational Chern-Simons term also and compare
with the results from the anomaly approach of Sec. III in order to
see whether they both agree or not. This would provide a non-trivial
check of the AdS/CFT-correspondence beyond the Einstein-Hilbert
action;
there are some works already in this direction
\cite{Cho:98,Blag:04,Blag:03}, but there are several aspects which
should be clarified.

There are two ``classically'' equivalent approaches for this
purpose. These are the purely gravity approach of Brown-Henneaux
\cite{Brow:86} and Chern-Simons (CS) approach. Here, let me consider
the latter approach since it is easier and provides some explicit
computations of the symmetry generators and their Dirac brackets of
(\ref{Virasoro:class}) even {\it far} from asymptotic boundary,
which are $not$ available in the former approach. Moreover, it can
reveal the holographic phenomena explicitly and the novel boundary
effects to the derivative of Dirac delta function, which are the
mathematical
origin of the classical central terms \cite{Park:98}.\\

\begin{subsection}
{Chern-Simons gauge theory with boundaries}
\end{subsection}

It is well known that CS (gauge) theory with boundaries produces
central terms in Virasoro algebras, as well as in Kac-Moody
algebras, even at ``classical'' level; this has been first spelled
out in \cite{Bana:99}, but rigorously computed later in
\cite{Park:98,Park:99}. This is a general field theoretic result
only if some appropriate boundary conditions are satisfied,
regardless of the physical contents of the CS theory.
Moreover, this is {\it not} an artifact of a ``classical'' theory,
but persists even in quantum theory because it can not be removed
from some quantum effects due to normal orderings \cite{Park:99}.

So, if a theory can be expressed as the CS theory with the
appropriate boundary conditions, one can quickly identify the
Kac-Moody and Virasoro algebras with the classical central terms.
This is actually the case of three-dimensional Einstein gravity with
a cosmological constant $\Lambda$, where the usual BTZ black hole or
the three-dimensional Kerr-de Sitter solutions ($KdS_3$) are
admitted, depending on the sign of $\La$ \cite{Bana:99,Park:98}.

The generalization of this approach to some more general class of
gravity systems, i.e., with matter couplings \cite{Park:04a} or with
higher curvature terms \cite{Said:00,Krau:05b,Saho:06} would not be
possible, in general. But, the three-dimensional gravity with a
gravitational Chern-Simons term is an exceptional case since the
gravitational Chern-Simons term itself can also be expressed as the
CS theory for another choice of the invariant quadratic forms of the
Lie algebra, for a non-vanishing $\La$ \cite{Witt:88}; on the other
hand, for the case of $\La=0$, the quadratic forms are not well
defined since they are degenerate. So, for the most general form of
the invariant quadratic forms which admit the new choice for the
gravitational Chern-Simons action as well, one can express the
Einstein gravity with the gravitational Chern-Simons term and
non-vanishing $\La$ as a CS gauge theory \cite{Witt:88,Cho:98}.

Moreover, in the GCS-BTZ black holes, there is no difference in the
metric form, though there are some shifts in the ADM mass and
angular momentum, and so there is no difference in the boundary
conditions for the corresponding CS theory; however, this would not
be valid generally for other non-trivial solutions in which there
are some important deformations of the metric itself.

Hence, all the previous results about the bare BTZ black hole can be
applied to the GCS-BTZ case also, from the general results of the
Kac-Moody and Virasoro algebras for the CS theory.\\

\begin{subsection}
{ SO(2,2) Chern-Simons gravity with the gravitational Chern-Simons
term}
\end{subsection}

For the (2+1)-dimensional space with a negative cosmological
constant $\La=-1/l^2$, symmetry of the space is ${SO(2,2)}$,
which has the following commutation relations among the generators
of the Lie group
\begin{\eq}
\label{Poin}
 [J_a, J_b ]={\ep_{ab}}^c J_c,~~ [J_a, P_b]={\ep_{ab}}^c P_c,
 ~~[P_a, P_b ]=\f{1}{l^2}{\ep_{ab}}^c J_c.
\end{\eq}
The most general form of the invariant quadratic forms are
\cite{Witt:88,Cho:98}
\begin{\eq}
\label{quad}
 \left< J_a, J_b \right>= \al \eta_{ab},~~\left< J_a, P_b
\right>= \be \eta_{ab},~~\left< P_a, P_b \right>= \f{\al}{l^2}
\eta_{ab}.
\end{\eq}
Here, $\al$ and $\be$ are some arbitrary constants, but the ratio of
$\left< J_a, J_b \right>$ and $\left< P_a, P_b \right>$ are
completely fixed by the algebras (\ref{Poin}).

The algebras (\ref{Poin}) and the quadratic forms (\ref{quad}) look
unusual. But, if I introduce
\begin{\eq}
J^\pm_a=\f{1}{2} (J_a \pm l P_a),
\end{\eq}
(\ref{Poin}) and (\ref{quad}) become
\begin{\eq}
\label{Poin_usual}
 && [J^{\pm}_a, J^{\pm}_b ]={\ep_{ab}}^c J^{\pm}_c,~~
 [J^{\pm}_a, J^{\pm}_b]=0 ,\\
&&\left< J^{\pm}_a, J^{\pm}_b \right>= \f{1}{2} (\al \pm \be l)
\eta_{ab},~~ \left< J^{\pm}_a, J^{\pm}_b \right>= 0.
\label{quad:usual}
\end{\eq}
This is the usual form of the ${SL(2,{\bf R})} \times { SL(2,{\bf
R})}$ Lie algebra but with different values of the quadratic forms
of the two sectors.

Now, by considering the Lie-algebra-valued one-form
\begin{\eq}
{\bf A} &=&\om^a J_a +e^a P_a =A^{+} +A^{-}, \no \\
A^{\pm}&=&(\om^a \pm \f{e^a}{l} )J^{\pm}_a
\end{\eq}
with the triads $e^a ={e^a}_{\mu} dx^{\mu}$ and the spin connections
$\om^a=(1/2)\ep^{abc} \om_{\mu bc}dx^{\mu}$ \footnote{ The
definition depends on the signature of the internal metric
$\eta_{ab}$. Our formulae are the case where the number of negative
signatures is odd. For more details about my conventions, see
Appendix {\bf A}.}, the CS action becomes [ $\left< A\we B\right>$
is understood as $\left< A  \we {,} B\right>$ ], up to some boundary
terms,
\begin{\eq}
\label{CSgravity}
 I_{CS}[\bf{A}]&=& \frac{k}{4 \pi} \int_{\cal M} \left<
{\bf A}
\wedge d {\bf A}+\frac{2}{3}~ {\bf A} \wedge {\bf  A} \wedge {\bf A}  \right> \nonumber \\
&=& \frac{k}{4 \pi}\Om^+ \int_{\cal M} Tr \left( A^{+}\wedge d
A^{+}+\frac{2}{3}~ A^{+} \wedge A^{+} \wedge A^{+} \right) -
~~(+\lra -)
\nonumber \\
&=&\frac{k \be}{\pi} \int_{\cal M} Tr \left( e \wedge {
R}+\frac{1}{3}~ e
\wedge e \wedge e  \right) \no \\
 &&+\frac{k \alpha}{2 \pi} \int_{\cal M} Tr \left( \omega \wedge \left(d
\omega+\frac{2}{3}~ \omega \wedge \omega \right) +\frac{e}{l^2}
\wedge { T} \right),
\end{\eq}
where $\Om^{\pm}=\be l \pm \al$, $Tr(J^{\pm}_a J^{\pm}_b)=(1/2)
\eta_{ab}$ and
\begin{\eq}
{R}&=&d \om +\om \wedge \om \no \\
&=&\f{1}{2}{R^a}_{b \nu \mu} dx^{\nu} \wedge dx^{\mu} \no \\
&=&\f{1}{2} {e^a}_{\al} {e_b}^{\be} {R^{\al}}_{\be \nu \mu}dx^{\nu}
\wedge
dx^{\mu}, \no \\
T&=&de +2 \omega \wedge e \no \\
&=&\f{1}{2}{T^a}_{\nu \mu}dx^{\nu} \wedge dx^{\mu} \no
\\
&=&\f{1}{2}{e^a}_{\al}{T^{\al}}_{\nu \mu}dx^{\nu} \wedge dx^{\mu}
\end{\eq}
are the curvature and torsion 2-forms, respectively.

The equations of motion of the CS gravity, by treating $A^{+}$ and
$A^{-}$ ``independently'', become the usual forms
\begin{\eq}
\label{F=0}
F^{\pm}&=&dA^{\pm} +A^{\pm} \wedge A^{\pm} \no \\
&=&R  +\f{1}{l^2} e \pm \f{1}{l} T\wedge e =0
\end{\eq}
or
\begin{\eq}
\label{T=0}
&&T=0, \\
\label{R=0}
 &&R +\f{1}{l^2} e \wedge e =0,
\end{\eq}
where I have chosen the boundary conditions
\cite{Bana:99,Oh:98,Park:99,Park:98}, for each time slice,
\begin{eqnarray}
\label{bc1}
&&A_0|_{\partial M} \propto A_{\varphi}|_{\partial M}, \\
\label{bc2}
 &&\oint _{\partial M} dt d \varphi \left< A_{\varphi},
A_{\varphi} \right> =\mbox{fixed},
\end{eqnarray}
with the boundary action
\begin{\eq}
I_S=-\f{k}{4 \pi} \oint_{\partial M} dt d \varphi \left<
A_{\varphi}, A_0 \right>.
\end{\eq}

Here, I note that the equivalence of the equations (\ref{F=0}) or
(\ref{T=0}, \ref{R=0}) and the Einstein equations (\ref{eom}) can be
achieved only after solving the torsion-free condition (\ref{T=0})
first. This should be the case since the spin-connections $\om$ are
not independent variables but are determined by the torsion-free
condition (\ref{T=0:comp}) already. Actually, by plugging
(\ref{T=0}) into the action (\ref{CSgravity}), it is a standard
computation to show that (\ref{CSgravity}) is equivalent to the
gravity action (\ref{EHGCS}), up to some boundary terms, with the
couplings (see Appendix {\bf A} for details)
\begin{\eq}
\label{connect}
 k \be=-\f{1}{4 G},~~\f{\al}{l \be} =\hb.
\end{\eq}

But at this point, there is one subtlety here: The whole CS
equations of motion are not available when one of $\Om^{\pm}$'s
vanishes and this occurs with $\be l=\al$ or $\hb=2$. In this
critical case I have only one sector of the solutions in
(\ref{F=0}), such as the torsion-free condition (\ref{T=0}) is not
``necessarily '' required. So, the equivalence of CS gravity
(\ref{CSgravity}) with the gravitational-Chern-Simons-corrected
gravity (\ref{EHGCS}) can $not$ be achieved in this case, generally.
However, if I restrict the solution space to the torsion-free ones
only, the equivalence is admitted still. This is actually the
situation that I consider in this paper since the BTZ solution
(\ref{BTZ}) satisfies (\ref{T=0}) and (\ref{R=0}), which do not
depend on the choice of $\om$ or $e$.

Now, in order to study the black hole solution (\ref{BTZ}), in the
context of the CS gravity, it is convenient to introduce a proper
radial coordinate $\rho$, such as (\ref{BTZ}) can be written as
\footnote{ Note that the
 sign convention of $\varphi$ differs from Ref. \cite{Bana:99}, such
 as it agrees with the original BTZ metric (\ref{BTZ})
 \cite{Bana:92}.
 This agrees also with Refs. \cite{Krau:05a,Solo:05a,Saho:06,Bala:99}. }
\begin{eqnarray}
ds^2 =-\mbox{sinh}^2 \rho \left( \frac{r_+ dt}{l} -r_{-} d {\varphi}
\right) ^2 +l^2 d \rho^2 +\mbox{cosh}^2 \rho \left( \frac{r_{-}
dt}{l} -r_+ d \varphi \right )^2
\end{eqnarray}
with
\begin{eqnarray}
r^2=r^2_+ \mbox{cosh}^2 \rho - r^2_{-} \mbox{sinh}^2 \rho.
\end{eqnarray}
In these coordinates, the (outer) event horizon is at $\rho =0$ and
hence this metric describes the exterior of the horizon for real
values of $\rho$, but the interior for imaginary values of $\rho$.
Then, it is easily checked that the 1-form gauge connections are
given, in the proper coordinates, by
\begin{eqnarray}
{{\bf A}^{\pm}}^0& =&\pm\frac{r_+ \pm r_{-}}{l} \left( \frac{dt}{l}
\mp d {\varphi}
\right) \sinh \rho , \nonumber \\
{{\bf A}^{\pm}}^{ 1} &=& \pm  d \rho, \nonumber \\
{{\bf A}^{\pm}}^{ 2} &=&\frac{r_+ \pm r _{-}}{l} \left( \frac{dt}{l}
\mp d \varphi \right) \cosh \rho.
\end{eqnarray}
[ The superscript indices denote the group indices $a=0,1,2$.] In
matrix form \footnote{I take $ J_0=\frac{1}{2}
\left(\begin{array}{cc} 0 & -1 \\ 1 & 0 \end{array} \right) ,~
J_1=\frac{1}{2} \left(\begin{array}{cc} 1 & 0 \\ 0 & -1 \end{array}
\right),~ J_2 =\frac{1}{2} \left( \begin{array}{cc} 0 & 1 \\ 1 & 0
\end{array}
 \right)$, and $\epsilon_{012}=1$ as in Ref. \cite{Bana:99}. The final results about
 the Virasoro algebras, however, do not depend on
 the choice of the representation.}, this becomes
\begin{eqnarray}
{\bf A}^{\pm} =\frac{1}{2} \left(
\begin{array}{cc}
 \pm d \rho & {z_{\pm}}~ e^{\mp \rho} dx^{\pm} \\
 z_{\pm}~ e^{\pm \rho} dx^{\pm} & \mp d \rho
\end{array}\right),
\end{eqnarray}
where ${z_{\pm}}\equiv(r_+ \mp r_{-})/l$ and $x^{\pm}=t/l \pm
\varphi$. From this, the polar components \footnote{Here,
$A_{\rho}=\hat{\rho}^i A_i, A_{\varphi} =\hat{\varphi}^i A_i$, for
the orthogonal unit vectors $\hat{\rho}, \hat{\varphi}$ on the
spatial boundary $\pa \Sigma$ with ${\cal M}=\Sigma \times R$; $\Si$
is a $2$-dimensional disc of space, and $R$ is a $1$-dimensional
infinite, real manifold of time.} in the proper coordinates can be
obtained as
\begin{eqnarray}
\label{A:BTZ}
 A^{\pm}_{\rho}=\pm J_1,~A^{\pm}_{\varphi}=\mp
{z_{\pm}} ~({U}^{-1} J_2 {U} ),~ A^{\pm}_t =\mp l A^{\pm}_{\varphi}
\end{eqnarray}
with
\begin{eqnarray}
{U} = \left(
\begin{array}{cc}
e^{\pm \rho /2} &0 \\
0 & e^{\mp\rho/2}
\end{array} \right).
\end{eqnarray}
Here, I note that this solution satisfies the boundary conditions
(\ref{bc1}) and (\ref{bc2}) for $any$ radius $\rho$, such as the
solution can be implemented even at the boundary whose radius may be
arbitrary, from  $0$ (at $r_+$) to $\infty$. And, the condition
(\ref{bc2}) implies the $micro-canonical$ ensemble, from the
relation $\left< A^{\pm}_{\varphi}, A^{\pm}_{\varphi} \right> \sim
(m \pm
j/l)$.\\

\begin{subsection}
{Symmetry algebras with classical central terms and statistical
entropy }
\end{subsection}

The CS action has
diffeomorphism ({\it Diff}) symmetries.
If there are boundaries, central terms appear in the symmetry
algebras, even at the classical level.
For a spatial and time-independent {\it Diff}:
\begin{eqnarray}
\delta_{f} x^{\mu} &=&-\delta^{\mu}_{~ k} {f^{\pm}}^{ k}, \nonumber \\
\delta_{f} {A_i^{\pm}}^{ a}  &=&{f^{\pm}}^{ k} \partial_k
{A_i^{\pm}}^{ a} +
(\partial_i {f^{\pm}}^{ k}) {A_k^{\pm}}^{ a}, \nonumber \\
\delta_{f} {A_0^{\pm}}^{ a} &=&{f^{\pm}}^{ k} \partial_k
{A_0^{\pm}}^{ a},
\end{eqnarray}
the Lagrangian of (\ref{CSgravity}) transforms as $\delta_{f}
L_{CS}=dX^{\pm}_f/dt$ with $X^{\pm}_f=(k \Om^{\pm})/(4 \pi) \oint
_{\Sigma} d^2x~
  Tr(
  {f^\pm}^{\rho} A_{\rho}^{\pm} A_{\varphi}^{\pm} )$
  when the boundary conditions ``${A^{\pm}}^{ a}_{\rho} |_{\partial
  \Si}$=constant'' is imposed, which is a quite natural choice according
  to the explicit BTZ solution
  (\ref{A:BTZ}).

Then, the conserved Noether charges become
\begin{eqnarray}
\label{Q:diff}
 Q ^{\pm}(f) &=&\f{k \Om^{\pm}}{4 \pi} \int_{\Sigma}
d^2 x ~
Tr({f^{\pm}}^{ k} A^{\pm}_k \epsilon^{ij} F^{\pm}_{ij} ) \no \\
&&- \f{k \Om^{\pm}}{4 \pi}  \oint_{\partial
  \Si} d \varphi ~Tr( 2 {f^{\pm}}^{ \rho} A^{\pm}_{\rho} A^{\pm}_{\varphi}
+{f^{\pm}}^{ \varphi} A^{\pm }_{\varphi}A^{\pm}_{\varphi}+
{f^{\pm}}^{ \varphi} A^{\pm }_{\rho}A^{\pm}_{\rho})  \\
  & \equiv & Q^{\pm}_{B} (f) + Q^{\pm}_{S} (f) \nonumber
\end{eqnarray}
with the bulk and boundary parts $Q_B^{\pm}(f)$ and $Q_S^{\pm}(f)$,
respectively;
the last constant term,
proportional to $Tr(A_{\rho} A_{\rho})$, in (\ref{Q:diff}) was
included to obtain the $standard$ Virasoro central term, with the
help of the ambiguities in the definition of Noether charge. These
satisfy the Virasoro algebras with $classical$ central terms in
Dirac bracket algebras
\begin{\eq}
\label{Virasoro:Q} \{ Q^{\pm}(f), Q^{\pm}(g) \}^* &\approx & \{
Q_S^{\pm}(f),
Q_S^{\pm}(g) \}^* \no \\
&\approx &Q_S([f,g]) -\f{k \Om^{\pm}}{2 \pi}~Tr(A^{\pm}_{\rho}
A^{\pm}_{\rho} ) \oint_{\partial \Si} d \varphi ( {f^{\pm}}^{ \rho}
\partial_{\varphi} {g^{\pm}}^{ \rho}- {f^{\pm}}^{ \varphi}
\partial_{\varphi} {g^{\pm}}^{ \varphi} ),
\end{\eq}
where $[f,g]^k=f^{\varphi} \partial _{\varphi} g^k -g^{\varphi}
\partial _{\varphi} f^k$ is Lie bracket on the boundary circle
($\partial \Si$).

Under the {\it Diff} generated by the Noether charges $Q^{\pm}(f)$,
the gauge fields of (\ref{A:BTZ}), representing the BTZ black hole,
have the transformations
\begin{eqnarray}
\delta_f A^{\pm}_{\varphi} &=&\frac{1}{2} \left(
    \begin{array}{cc}
\pm \partial_{\varphi} {f^{\pm}}^{\rho} & z_{\pm}~ e^{\mp\rho}
({f^{\pm}}^{\rho} \mp
\partial_{\varphi}
{f^{\pm}}^{\varphi}) \\
-z_{\pm}~ e^{\pm \rho} ({f^{\pm}}^{\rho} \pm \partial_{\varphi}
{f^{\pm}}^{\varphi})
& \mp \partial_{\varphi} {f^{\pm}}^{\rho} \end{array} \right), \no \\
\delta_f A^{\pm}_{\rho} &=&0.
\end{eqnarray}
This implies  that the black hole solution (\ref{A:BTZ}) admits the
isometries, i.e., $\delta_f A^{\pm}_i=0$ as $\rho \ra \infty$ when
\begin{eqnarray}
\label{f:cond}
 {f^{\pm}}^{\rho}|_{\partial \Si}=-\partial_{\varphi}
{f^{\pm}}^{\varphi}|_{\partial \Si}
\end{eqnarray}
is satisfied, though $not$ necessarily for arbitrary $\rho$. This
exactly agrees, to the leading order, with the asymptotic isometries
found by Brown-Henneaux \cite{Brow:86}\footnote{ There are several
other ways to implement the {\it Diff} even for the finite values of
$\rho$ \cite{Carl:98}, where there are some $RG-flows$ of the
central charges and conformal weights without changing the
statistical entropies. So, there remains the question on the very
place where the black hole's degrees of freedom live.}. Contrary to
the existence of the central term itself, this result is a purely
non-Abelian effect which comes from the off-diagonal parts.

Now, by substituting (\ref{f:cond}) with the insertion of $Tr(
A^{\pm}_{\rho} A^{\pm}_{\rho})=1/2$ for the black hole solution
(\ref{A:BTZ}), the algebras (\ref{Virasoro:Q}) become the standard
Virasoro algebras, in the coordinate space, with $classical$ central
charges
\begin{eqnarray}
\label{c:class}
 c^{\pm}=-12 k \Om^{\pm}Tr( A^{\pm}_{\rho}
A^{\pm}_{\rho} ) =\ga^{\pm}\f{3l}{2G}
\end{eqnarray}
with $\ga^{\pm}=1 \pm \hb$. In the $\hb \ra 0$ limit,
  these classical central charges reduce to the usual result of Brown-Henneaux \cite{Brow:86} for the
 asymptotic isometry of $AdS_3$  and also agrees exactly with that
 of conformal anomaly computation \cite{Henn:98,Hyun:99,Bala:99}. But
 interestingly, the $\hb$-dependent central charges give an exact
 agreement also with the $semi-classical$ central charges
\begin{\eq}
 \hat{c}^{\pm}=\f{c^{\pm}}{\hbar},
\end{\eq}
 as in (\ref{c-c}),
 that have been obtained from gravitational anomaly computation; this
 seems to be a quite non-trivial result since I don't see any general proof
 about the equivalence of the two central charges even without the
 gravitational Chern-Simons term
 though it seems to be quite plausible in the context of AdS/CFT
 correspondence, which identifies the ``classical'' asymptotic CFT of AdS space
 on the one hand with the ``quantum-mechanical'' CFT on the boundary on the other hand.

The more familiar
 momentum-space Virasoro algebras (\ref{Virasoro:class}) can be obtained by
 defining the boundary parts of the Noether charges in
(\ref{Q:diff}) as
\begin{\eq}
Q^{\pm}_S(f) \equiv \frac{1}{2 \pi} \oint _{\partial \Si} d \varphi
{f^{\pm}}^{\varphi}
 \left(\sum_n L^{\pm}_n e^{+in \varphi}\right)
\end{\eq}
and the central charges are given by (\ref{c:class}). The ground
state generators, from the definition, become
\begin{eqnarray}
\label{ground:class}
 L_0^{\pm} =- \frac{k \Om^{\pm}}{4\pi} \oint_{\partial \Si} d \varphi~ Tr( A^{\pm}_{\varphi} A^{\pm}_{\varphi}
 +A^{\pm}_{\rho} A^{\pm}_{\rho} ) = \ga^{\pm} \f{1}{2}(l m \pm j)+
 \f{c^{\pm}}{24}.
\end{eqnarray}

Note that the $\hb$-dependent terms, as well as $\hb$-independent
terms, agree exactly with $\hat{L}^{\pm}_0=L^{\pm}/\hbar$ of
(\ref{ground}). So, if I define the black hole's mass and angular
momentum {\it canonically} as in (\ref{ground}), from the general
consideration of CFT on the torus \cite{Fran:97},
\begin{\eq}
L^{\pm}_0=\f{l M \pm J}{2} + \f{c^{\pm}}{24}
\end{\eq}
one obtains the same mass and angular momentum as in the anomaly
approach \cite{Krau:05a,Solo:05a}, which agree with the usual ADM
quantities of (\ref{M_J}) \cite{Garc:03,Dese:05}
also. It
does not seem that this is not just a coincidence but there be some
deep reasons  involving the {\it holographic} principle; however,
our CFT computation of the statistical entropy does not depend on
the manners of identifications of $M$ and $J$ but only on the
geometrical quantities of $r_+$ and $r_-$, such as the CFT
computation provides a quite independent estimation of the {\it
would-be} black hole entropy.

Now with the Virasoro algebras with ``classical'' data of the
central charges (\ref{c:class}) and the ground state generator
$L^{\pm}_0$ in (\ref{ground:class}), it is straightforward to obtain
the corresponding quantum Virasoro algebras \cite{Park:02}: If I
consider the canonical quantization
\begin{eqnarray}
[{\bf L}_m^{\pm}, {\bf L}_n^{\pm} ]=i \hbar \{L_m^{\pm}, L_n^{\pm}
\}^{*}
\end{eqnarray}
for the quantum operators ${\bf L}_m^{\pm}$ and a rescaling
transformation
\begin{eqnarray}
{\bf L}^{\pm}_m \rightarrow \hbar (: \hat{L}^{\pm}_m: +\hbar a^{\pm}
\delta_{m,0} )
\end{eqnarray}
for the normal ordered operators $:\hat{L}_m^{\pm}:$ with some
possible normal ordering constants $a^{\pm}$, one can easily find
the corresponding quantum Virasoro algebras
\begin{eqnarray}
[:\hat{L}_m^{\pm}:,:\hat{L}_n^{\pm}:]=(m-n) :\hat{L}_{m+n}^{\pm}:
+\frac{\hat{c}^{\pm}_{\hbox{\scriptsize tot}}}{12} m(m^2-1)
\delta_{m,-n}
\end{eqnarray}
with
\begin{eqnarray}
\label{ctot} \hat{c}^{\pm}_{\hbox{\scriptsize
tot}}=\frac{c^{\pm}}{\hbar} +{c}^{\pm}_{\hbox{\scriptsize quant}}.
\end{eqnarray}
Here, the quantum correction $c^{\pm}_{\hbox{\scriptsize quant}}$ is
due to some operator re-orderings and it is order of $O(1)$.

With the Virasoro algebras of $:\hat{L}_m:$ in the standard form,
which is defined on the {plane}, one can use the Cardy formula for
the asymptotic states \cite{Park:02,Kang:04,Park:04b} as in
(\ref{Cardy})
\begin{eqnarray}
\mbox{log} \rho (\hat{\Delta}^{\pm}) \simeq 2 \pi \sqrt{ \f{1}{6}
\left(\hat{c}^{\pm}_{\hbox{\scriptsize tot}}-24
  \hat{\Delta}^{\pm}_{\hbox{\scriptsize
  min}}\right)\left(\hat{\Delta}^{\pm}-\frac{\hat{c}^{\pm}_{\hbox{\scriptsize tot}}}{24}\right) },
\end{eqnarray}
where $\hat{\Delta}^{\pm}$ are the eigenvalues of
$:\hat{L}^{\pm}_0:$ for the black-hole quantum states
$|\hat{\Delta}^{\pm}\rangle$, and
$\hat{\Delta}^{\pm}_{\hbox{\scriptsize min}}$ are their minimum
values. When expressed in terms of the classical generators
$L_0^{\pm}$ and the central charges $c^{\pm}$ through
\begin{eqnarray}
\hat{\Delta}^{\pm}=\f{L^{\pm}_0}{\hbar}- \hbar a^{\pm},
\end{eqnarray}
one obtains
\begin{eqnarray}
\mbox{log} \rho (L^{\pm}_0) \simeq \frac{2 \pi}{\hbar} \sqrt{
 \f{1}{6} \left(c^{\pm}-24
 L^{\pm}_{0 ~\hbox{\scriptsize min}}+\hbar c^{\pm}_{\hbox{\scriptsize quant}}+24 \hbar^2
 a^{\pm}\right)\left(L^{\pm}_0-\frac{c^{\pm}}{24}-\frac{\hbar
 c^{\pm}_{\hbox{\scriptsize quant}}}{24}-\hbar^2 a^{\pm} \right) }.
\end{eqnarray}
This approach shows explicitly how the {\it classical} Virasoro
generators $L_0^{\pm}$ and central charges $c^{\pm}$ can give the
correct order of the semiclassical Bekenstein-Hawking entropy
(\ref{S:BH}),
\begin{eqnarray}
\label{BH} S_{\hbox{\scriptsize BH}}\simeq\frac{{\cal A}_+}{4 \hbar
G}
\end{eqnarray}
with $\sqrt{c^{\pm} L_0^{\pm}} \sim {\cal A}_+/G$; the quantum
corrections due to reordering give the negligible order of $O(1)$
effect to the entropy when one considers the large black holes with
${\cal A}_+/(G \hbar) \gg 1$.

Then, the statistical entropy for the asymptotic states becomes [
omitting the small quantum corrections of the order of $O(1)$ ]
\begin{\eq}
S_{\stat}&=&\mbox{log}~ \rho ({L}^{+}_0)+\mbox{log}~ \rho
({L}^{-}_0) \no \\
&=&\f{\pi}{4 G \hbar} (|\ga^+| +|\ga^-|) r_+ +\f{\pi}{4 G \hbar}
(|\ga^+|-|\ga^-|)r_-~, \label{S:stat(class)}
\end{\eq}
where I have chosen ${L}^{\pm}_{0~ (\hbox{\scriptsize min})}=0$,
which corresponds to the $AdS_3$ vacuum solution with $m=-1/(8G)$
and $ j=0$, in agreement with (\ref{S:stat}). This has exact
matchings with (\ref{S:stat}) in the $\hb$-dependent correction
terms, as well as $\hb$-independent terms. I note also that the
``$1/\hbar$''-factor in the black hole entropy (\ref{S:stat(class)})
was generated in the process of canonical quantization of the
classical Virasoro algebras.

So, the statistical entropy, based on the classical symmetry
algebras, agrees with the thermodynamic black hole entropy even in
the correction terms due to the gravitational Chern-Simons term, as
well as the usual one for the Einstein-Hilbert action.
This might a subtle issue because of some normalization differences
between the different bases and conventions in the literatures.
Actually, there are ubiquitous factor ``2'' differences between
different bases. So, I have included some details about the
transformations of the formulae between the different bases and
conventions in the Appendix {\bf A} in order to ensure that this
exact factor matching is a {\it
solid} result, actually.\\

\begin{section}
{Summary and discussions}
\end{section}

I have studied the thermodynamics of BTZ black hole in the presence
of the higher-derivative corrections of the gravitational
Chern-Simons term and its solid connection with the statistical
approaches, based on the holographic anomalies and the classical
symmetry algebras.

The main results are as follows:

First, for the case of large coupling $|\hb| >1$ the new entropy
formula
is proposed from the purely thermodynamic
point of view such as the second law of thermodynamics be
guaranteed\footnote{This has been proved more rigorously and
extensively in Ref. \cite{Park:0611}.}.

Second, I have found
supports of the proposal from the CFT
based approaches which reproduce the new entropy formulae for
$|\hb|>1$, as well as the usual entropy formula for the small
coupling case of $|\hb| \le 1$.

Third, I have found the exact ``factor'' matchings between the
holographic anomaly approach
and the classical symmetry algebra approach from the Chern-Simons
formulation of the three-dimensional gravity. This would provide a
non-trivial check of the AdS/CFT-correspondence in the presence of
higher-derivative terms in the gravity action.


Now, several comments are in order.

1. {\it On the general validity of Cardy formula with
higher-derivative/curvature corrections }: It is interesting to note
that the statistical entropy (\ref{S:stat}) from the Cardy formula
(\ref{Cardy}) has basically the {\it same form} for both the
Einstein-Hilbert action and the gravitational-Chern-Simons-corrected
action; {\it the only changes are some correction terms in the
central charges and conformal weights themselves, rather than
considering the higher-order corrections to the Cardy formula as in
Ref. \cite{Park:04b}.}
 This seems to be true even in the presence of
higher-``curvature'' terms \cite{Said:00,Krau:05b,Saho:06} and also
in the supersymmetric black holes \cite{Moha:05}. So, there should
exist some explanations about this and actually this is the case.
This comes from the fact that the higher-derivative/curvature
actions do {\it not} necessarily imply the quantum corrections
though the converse can be true \cite{tHoo:74}.
So,
if the higher-derivative/curvature gravities are treated
semiclassically by neglecting the back-reaction effects, which are
quantum effects, such as (\ref{cond2}) or (\ref{cond4}) is
satisfied, the saddle- point approximation for the Cardy formula
(\ref{Cardy}) and so the entropy formula (\ref{S:stat}) are good
approximations, even with the higher-derivative/curvature terms in
the gravity action \cite{Park:04b}. There is another factor whose
departure from unity is order of $O[\mbox{exp}\{-2 \pi
\hat{\De}^{\pm}_{\eff}(\hat{\De}^{\pm}
-\hat{\De}^{\pm}_{\min})/\hat{c}_{\eff}\}]$, but this correction, if
there is, is not comparable with the leading term (\ref{S:stat}) and
other higher-order corrections, by departing the semi-classical
limit of (\ref{cond4}); in our case of the GCS-BTZ black holes there
is already the corrections of order of $O(r_-/r_+)$ in the leading
entropy (\ref{S:stat}), but this dominates the exponentially
suppressed corrections. Hence, {\it the leading Cardy formula
(\ref{Cardy}) would have quite general validity for any kinds of
semiclassical black holes in the higher-derivative/curvature
gravities unless the condition (\ref{cond1}) or (\ref{cond3}) is
violated}.

2. {\it Higher-order corrections to the saddle-point approximation}:
By relaxing the semiclassical condition of (\ref{cond2}) or
(\ref{cond4}) but keeping only the condition (\ref{cond1}) or
(\ref{cond3}), the higher order corrections in the Cardy formula
(\ref{Cardy}) can be evaluated by the steepest descent method, known
as the Rademacher expansion \cite{Rade:38,Dijk:00}. The statistical
entropy then becomes, up to fourth order, (see Appendix {\bf B} for
the details) from (\ref{SCFT4})
\begin{\eq}
\label{Cardy:higher}
 S_{\hbox{\scriptsize stat}(4)}&=&(S^+_0 + S^-_0)-\f{3}{2} \log
(S^+_0 S^-_0) + \log(\hat{c}^+_{\eff} \hat{c}^-_{\eff} )+ \log
(\pi^3/18)-\f{3}{8} \left(\f{1}{S^+_0} + \f{1}{S^-_0}
\right)+O((S^{\pm})^{-2} ) \no \\
&=&S_{\stat}-\f{3}{2} \log \left(\left(\f{\pi}{G \hbar} \right)^2
|\ga^+ \ga^- |(r^2_+ -r^2_-) \right) + \log \left(\ga^+ \ga^-
\left(\f{3l}{2 G \hbar} \right)^2 \right) + \log (\pi^3/18) \no
\\
&&-\f{3}{8} \left(\f{G \hbar}{\pi} \right)^2 \f{S_{\stat}}{|\ga^+
\ga^- |(r_+^2-r_-^2)}+O((G \hbar)^2/r^2_+,~ (G \hbar)^2/r^2_- ),
\end{\eq}
where $S^\pm_0$ denote the right/left-moving parts of the leading
entropy formula (\ref{S:stat}), i.e., $S^\pm_0=\log
\rho(\hat{\De}^{\pm})$ with $S^+_0 + S^-_0=S_{\stat}$ and this is
the expansion about the Planck constant $\hbar$. It would be a
challenging problem to compute the loop-corrected black hole
entropies in the gravity side also and compare with the above CFT
result (\ref{Cardy:higher}). Actually, the loop corrections in the
gravity side would not be trivial in this case since there would be
now some propagating mode(s) with the gravitational Chern-Simons
term \cite{Dese:82,Dese:91,
Miel:03}, in contrast to the
usual BTZ black hole \cite{Park:04b,Carl:97}.

$ 2\f{1}{2}$. {\it Subtleties of extremal and near-extremal black
holes }: If I consider extremal bare black holes with $r_+=r_-$,
i.e., $\hat{\De}^-_{\eff}=0$, which saturates the mass bound $m=j/l$
and has vanishing temperatures, there seem to exist some subtleties
in the above manipulations. Namely, the condition (\ref{cond4}) does
not apply and the back-reaction effect would not be negligible
anymore in this case, such as I would need to consider ``infinite''
higher-order corrections in the steepest-descent approximations,
which seems to be highly divergent from (\ref{Cardy:higher}); other
infinite series of exponential correction terms are actually of the
form $O[(\hat{\De}^{\pm}-\hat{\De}^{\pm}_{\min})^m
(\hat{\De}^{\pm}_{\eff}/\hat{c}_{\eff})^n \mbox{exp}\{-2 \pi
\hat{\De}^{\pm}_{\eff}(\hat{\De}^{\pm}
-\hat{\De}^{\pm}_{\min})/\hat{c}_{\eff}\}]$ with some positive
integers $m$ and $n$ \cite{Park:04b} such as the problematic part
does not contribute further. But, actually this is not quite correct
as can be seen easily in the original partition function
(\ref{CFTZ}). In the case of extremal bare black holes, the
left-moving sector is absent in the partition function because of
$\hat{L}_0^--\hat{c}^-/24=0$ such as total partition function is
given by, from (\ref{SCFT4}),
\begin{\eq}
\label{Cardy:extrem}
 S_{\hbox{\scriptsize stat(4): extreme}}
&=&\f{2 \pi r_+}{4 G \hbar} |\ga^+|-\f{3}{2} \log \left(\f{2\pi
r_+}{4G \hbar} |\ga^+| \right) + \log \left(\ga^+ \f{3l}{2 G \hbar}
\right) + \f{1}{2}\log (\pi^3/18) \no
\\
&&-\f{3}{8} \left(\f{4G \hbar}{2 \pi r_+} \right)
\f{1}{|\ga^+|}+O((G \hbar)^2/r^2_+).
\end{\eq}
This gives the correct Bekenstein-Hawking entropy for the leading
term as can be also read from (\ref{Cardy:higher}) and there is no
divergence in each order\footnote{Interestingly, the factor
``$3/2$'' in the logarithmic term agrees with the corresponding
corrections in the induced $WZW$ model at the horizon within the
context of CS gravity, in contrast to the factor ``2'' mismatches in
the non-extremal black holes \cite{Park:04b}. But, it is subtle to
compare this with the purely gravity manipulation since there is no
clear way to resolve a similar divergence problem. }. This implies
that, in the ``near-extremal'' case, the naive divergence in each
term of (\ref{Cardy:higher}) would cancel each other and one would
have only some finite entropy. Actually, this seems to be supported
also by the exact Raedmacher expansion which shows that the {\it
exact} entropy with all higher-order corrections is bounded by, up
to some exponentially suppressed terms, the Bekenstein-Hawking
entropy, i.e., $0 \leq S_{\exact} < S_{BH}$ \cite{Birm:00}; if there
are no cancelations, the exact entropy $S_{\exact}$ would easily
violate the above Birmingham-Sen's bound. On the other hand, it is
important to note that the condition for the right-moving sector
$only$ can be satisfied, though not possible for the left-moving
sector, such as the extremal bare black hole with a vanishing
temperature does {\it not} always imply the necessity of the
higher-order corrections; however, its relevance to the
back-reaction effect is not clear \cite{Pres:91}.

 On the other hand, the case of critical coupling $|\hb|=1$, which
has the extremal bound $M^2=J^2/l^2$ but a {\it non-vanishing}
temperature, has similar subtleties. In this case, one of
$\ga^{\pm}$ vanishes such as the condition (\ref{cond1}) would be
ambiguous, even though overall $\ga^{\pm}$ factor can be canceled
for a non-vanishing $\ga^{\pm}$. And, the condition (\ref{cond2})
can not be satisfied either such as its entropy has similar
divergence problem from (\ref{Cardy:higher}), as in the bare
extremal black holes. The resolution is similar to the bare-extremal
black hole, and the appropriate statistical entropies are given by,
from (\ref{SCFT4}),
\begin{\eq}
\label{Cardy:extreme2}
 S_{\hbox{\scriptsize stat(4)}: ~\hb=\pm 1}
&=&\f{2 \pi (r_+ \pm r_-)}{4 G \hbar}-\f{3}{2} \log \left(\f{2 \pi
(r_+ +r_-) }{4G \hbar} \right) + \log \left(2 \f{3l}{2 G \hbar}
\right) + \f{1}{2}\log (\pi^3/18) \no
\\
&&-\f{3}{8} \left(\f{4G \hbar}{2\pi(r_+ + r_-)}\right) +O((G
\hbar)^2/(r^2_+ \pm r_-)^2)
\end{\eq}
for $\hb=\pm 1$ and these agree with the entropies
(\ref{S_stat:new3}, \ref{S_stat:new4}) in the leading order. But, if
I consider the extremal bare black holes further with $r_+=r_-$, the
entropy for $\hb=1$ case reduces to (\ref{Cardy:extrem}), whereas
that for $\hb=-1$ case has divergent higher order terms with the
vanishing entropy in the leading term. This subtleties can be
resolved again in the original partition function language; there,
the right-moving sector is absent, i.e., $L_0^+-c^+/24=0$ due to
$\ga^+=1$, whereas the left-moving sector is also absent, i.e.,
$L_0^--c^-/24=0$ due to $r_+=r_-$ such as one has only a {\it single
ground} state with $\rho(\hat{\De}^+,\hat{\De}^-)=1$; this system
satisfies the Nernst formulation of the third law of thermodynamics
\cite{Hawk:94},
i.e., $S_{\stat}=\log \rho=0$, to
{\it all orders} !

 3. {\it Probing inside the outer
horizon by the gravitational Chern-Simons action ?}: Although there
are some solid supports from the second law of thermodynamics and
the CFT approaches, the inner horizon's data, which are required in
the complete formulae, look strange still; of course, the necessity
of the inner horizon's data seems to be a quite general feature with
quantum corrections from the result of (\ref{Cardy:higher}), but the
problem is that it occurs even at the leading, classical level.
Actually, this would be much strange in the Euclidean method of
conical singularity \cite{Solo:05a} or in the Wald's approach to
compute the black hole entropy, which gives the same entropy formula
with the inner-horizon term even though it is given by some
integrals over the outer horizon \cite{Solo:05a,Blag:06}.
So, understanding the roles of the inner horizon's data appearing in
the \ther relations would be a challenging problem;
some possible probing, in the context of the
AdS/CFT, beyond the event horizon have been considered recently
\cite{Stei:94,Bala:04,Maed:06}, but this need further studies.

4. {\it Classical (in)stability of the $|\hb|>1$ black holes}: For
the large coupling of $|\hb|>1$, the black-hole angular momentum is
greater than its mass $J^2/l^2 \geq M^2$, and there are three known
cases which show this ``exotic'' property, including the
gravitational Chern-Simons case, in $D=3$ and $5$
\cite{Carl:91,
Park:06}. There are no similar black hole solutions in $D=2$ and
$4$, as far as I know. In $D\geq 6$, the ``ultra-spinning'' black
holes are possible in Einstein gravity \cite{Myer:86}, but it seems
that there is a {\it classical instability} under small
perturbations \cite{Empa:03}.
So, it would be interesting
to investigate this classical (in)stability in our exotic cases
also; there might exist some topological reasons for this, but it is
not clear in our case since there are propagating modes also
\cite{Miel:03}, in contrast to the ordinary BTZ black hole and
$KdS_3$ solution \cite{Park:98}.

5. {\it The first law of thermodynamics vs. Hawking radiation for
$|\hb|>1$}: A difficult problem of the new entropy formula for the
case of large coupling $|\hb|>1$ is that
it requires rather unusual characteristic temperature $T_-=\kappa/(2
\pi)|_{r_-}$, which is negative-valued (for $\hb>1$), or
${T_-}'=-T_-$ (for $\hb <-1$), and angular velocity $\Om_-$, which
is the inner-horizon angular velocity in BTZ if I ``assume'' the
first law of thermodynamics. The negative-valued temperature {\it
might} be understood from the existence of the upper bound of mass
in (\ref{M_bound}), i.e., $M \leq J/l$ with positive $M$ and $J$, as
in the spin-systems \cite{Kitt:67}. However, the {\it very} meanings
of $T_-, T_{-}'$ and $\Om_-$ in the Hawking radiation are not clear
since
%
the radiation spectrum is determined by the metric alone, in the
standard analysis initiated by Hawking \ci{Hawk:75}. So, {\it does
this work imply that two black holes with identical BTZ metrics will
emit radiation with different spectra, one a black body spectrum
corresponding to a positive temperature $T_+$ for the weak coupling
of $|\hb| \le 1$ and one a very non-black-body spectrum
corresponding to a negative temperature $T_-$ for the strong
coupling of $|\hb|>1$ ?}  Or, {\it does this imply that the first
law of thermodynamics is ``not'' satisfied for $|\hb|>1$ ? }

This would be a difficult question whose complete answer is still
missing. But here, I would like to only mention the possible
limitation of the standard approach in this system and how this
$might$ be circumvented. An important point for this is that
dynamical geometry responds differently under the emission of
Hawking radiation. For example, the emission of energy $\om$ would
reduce the black holes's mass $M$ from the conservation of energy,
but this corresponds to the change of the angular momentum $j$, as
well as $m$, in the ordinary BTZ black hole context, due to the
mixing of the mass and angular momentum as in (\ref{M_J}). This is
in sharp contrast to the case of ordinary black hole. This seems to
be a key point to understand the
Hawking radiation in our
system, and in this argument the conservations of energy and angular
momentum, which are not well enforced in the standard computation,
would have a crucial role. So, in this respect, the Parikh and
Wilczek's approach \ci{ Pari:00}, which provides a direct derivation
of Hawking radiation as a quantum tunneling by considering the
global conservation law naturally, would be an appropriate framework
to study the problem.
Currently this is under study.

6. {\it Green's function method and thermal equilibrium}: The
Green's function methods for determining the temperature of a black
hole require an equilibrium with matter at the corresponding
temperature \ci{Hart:76}.  This work now implies that the analysis,
for the strong coupling of $\hb
>1$, assumes such an equilibrium with `` some exotic surrounding
matter '' which has a negative temperature\footnote{The
determination of the equilibrium temperature from the `` fundamental
period '' in the thermal Green's function, known as the KMS (
Kugo-Martin-Schwinger ) condition \ci{Kubo:57},
can be defined without the implicit assumption of a positive
temperature, though not quite well-known in the gravity community (
see Ref. \ci{Brat:76}, for example ).}, with an upper bound of
energy levels as in spin systems: Otherwise, i.e., with the ordinary
surrounding matter, the negative temperature black hole can not be
at equilibrium with positive temperature surroundings since an
object with a negative temperature is hotter than one with any
positive temperature.

I suspect that this would be rather easy to understand in our case
from the fact that in the AdS  space the artificial container is not
needed in order to study the canonical ( or grand-canonical )
ensemble \ci{Hawk:83,Hawk:99}. But, in the context without the
explicit container, there is a critical angular velocity
\ci{Hawk:99} at which the action of the black hole or the partition
function of its corresponding CFT  diverges. However, I note that
the critical value is the same as the lower bound of $\Omega_{-}$,
such as we are beyond the critical point with our angular velocity
$\Omega_{-}$. So, from this fact, it seems clear that the ensemble,
if there is, in this strong coupling regime would be quite different
from that of the usual BTZ black hole such as one can not simply
apply the usual result to the strong coupling case. It seems that we
need an independent analysis for this case.
%
\\

\appendix

\begin{section}
{Conventions and some useful formulae in differential forms}
\end{section}

In this appendix, I summarize the conventions and some useful
formulae in differential forms used in this paper. I have also
included some details about the computations in order to ensure that
the $exact$ factor matching, which is directly related to the
relation in (\ref{connect}) is a quite solid result, regardless of
some normalization differences between different bases. I have used
the Lorentzian metric for the internal Lorentz indices
$\eta_{ab}=diag(-1,1,1)$ and $\ep_{012}=-\ep^{012}=1$. [ For the
$s$-negative signatures in the metric generally, a number of
formulae will contain the factor of $(-1)^s$
\cite{Horn:89,Miel:03,Wald:84}. ]

The invariant quadratic forms for the $SL(2, {\bf R})$ generators
are (\ref{quad:usual}), i.e.,
\begin{\eq}
\label{quad:pm} \left< J^{\pm}_a, J^{\pm}_b \right>= \f{1}{2}~ (\al
\pm \be l) \eta_{ab},~~ \left< J^{\pm}_a, J_b^{\mp} \right>= 0,
\end{\eq}
and the Lorentz indices are raised and lowered by the metric
$\eta_{ab}$. [ One can consider conveniently the invariant form as
$\left< J^{\pm}_a, J^{\pm}_b \right>= (\al \pm \be l) Tr(J^{\pm}_a
J^{\pm}_b)$ by considering the explicit matrix representation of the
generators with $Tr(J^{\pm}_a J^{\pm}_b)=(1/2)\eta_{ab}$ as in the
Sec. IV, but the final results do not depend on the representations;
thus, I will keep (\ref{quad:pm}) in this appendix. ]

Now let me prove (\ref{CSgravity}), (\ref{F=0}), and the relations
in (\ref{connect}). To this end, I first note that the CS action in
(\ref{CSgravity}) can be written, in the component form for the
internal space, as
\begin{\eq}
\label{CSgravity:comp}
 \f{4 \pi}{k} I_{CS}[{\bf A}]
= \int_{\cal M} \f{1}{2}~ \Om^+ \left( \eta_{ab} {A^{+}}^{a}\wedge d
{A^{+}}^{b}+\frac{1}{3} ~\ep_{abc} {A^{+}}^{a} \wedge {A^{+}}^{b}
\wedge {A^{+}}^{c} \right)~ -~\left(+\lra -\right),
\end{\eq}
where I have used
\begin{\eq}
\left< {\bf A} \wedge {\bf  A} \wedge {\bf A} \right> &=& \left<
{A^{+}} \wedge \f{1}{2} [{ A^{+}}, {A^{+}}] \right>
~+~(+\lra -) \no \\
&=& \f{1}{2} {A^{+}}^{a} \wedge {A^{+}}^{b} \wedge { A^{+}}^{d}
{\ep_{ab}}^c
\left<J^+_d , J^+_c  \right> +(+\lra -) \no \\
&=& \f{1}{2} {A^{+}}^{a} \wedge {A^{+}}^{b} \wedge { A^{+}}^{c}
\cdot \f{1}{2} \ep_{abc} (\al +\be l)+\f{1}{2} {A^{-}}^{a} \wedge
{A^{-}}^{b} \wedge { A^{-}}^{c} \cdot \f{1}{2} \ep_{abc} (\al -\be
l)
 \no \\
&=&\f{1}{4} \Om^+ \ep_{abc} {A^{+}}^{a} \wedge {A^{+}}^{b} \wedge
{A^{+}}^{c} -\f{1}{4} \Om^- \ep_{abc} {A^{-}}^{a} \wedge {A^{-}}^{b}
\wedge {A^{-}}^{c}
\end{\eq}
and
\begin{\eq}
\left< {\bf A}\wedge d {\bf A}\right>
&=& \left< {A^{+}}\wedge d {A^{+}}\right> +(+\lra -) \no \\
&=& {A^{+}}^{a}\wedge d {A^{+}}^{b}\left< J^+_a , J^+_b   \right> +(+\lra -) \no \\
&=& {A^{+}}^{a}\wedge d {A^{+}}^{b} \cdot \f{1}{2} (\al +\be
l)\eta_{ab}+ {A^{-}}^{a}\wedge d {A^{-}}^{b} \cdot \f{1}{2} (\al
-\be l)\eta_{ab} \no
\\
&=&\f{1}{2} \Om^+ \eta_{ab} {A^{+}}^{a}\wedge d {A^{+}}^{b}-\f{1}{2}
\Om^- \eta_{ab} {A^{-}}^{a}\wedge d {A^{-}}^{b},
\end{\eq}
with the one-form gauge fields ${\bf A}={A^{+}}^{a}
J^+_a+{A^{-}}^{a} J^-_a$ and $\Om^{\pm}=\be l \pm \al$.

By considering ${A^{\pm}}^{ a}=\om^a \pm  e^a/l$ with the spin
connections $\om^a$ and the triads $e^a$, one can find that the CS
action (\ref{CSgravity:comp}) becomes, after some manipulations,
\begin{\eq}
\label{CSgravity:comp2}
 \f{4 \pi}{k} I_{CS}
&=& \int_{\cal M} \left[ \alpha \omega^a \wedge \left(d
\omega_a+\frac{1}{3}~\ep_{abc} \omega^b \wedge \omega^c \right)
+\frac{\alpha}{l^2} e^a \wedge ( d e_a +\ep_{abc} \om^b \we e^c)\right. \no \\
&&\left.+\be e^a \wedge \left(2 d\omega_a+\ep_{abc} \omega^b \wedge
\omega^c +\frac{1}{3 l^2} ~\ep_{abc}  e^b \wedge e^c \right) -\be
d(\om^a \wedge
e_a ) \right]\no \\
 &=&\alpha \int_{\cal M} \left[
\omega^a \wedge \left(d \omega_a+\frac{1}{3}~\ep_{abc} \omega^b
\wedge \omega^c \right)
+\frac{1}{l^2} e^a \wedge { T_a}\right] \no \\
&&+\be \int_{\cal M} \left( 2 e^a \wedge { R_a}+\frac{1}{3 l^2}~
\ep_{abc} e^a \wedge e^b \wedge e^c  \right) -\be \oint_{\pa {\cal
M}} \om^a \wedge e_a ,
\end{\eq}
where I have defined the curvature two-form, in {\it vector} form
basis,
\begin{\eq} \label{R^a}
R^a&=&\f{1}{2}~ {\ep^a}_{bc} R^{bc} \no \\
  &=&d\omega_a+\frac{1}{2}~\ep_{abc} \omega^b \wedge \omega^c
\end{\eq}
from $R^{ab}=d \om^{ab}+{\om^a}_c \wedge \om^{cb}$ and
$\om_{ab}=-\ep_{abc} \om^c,~\om^a=(1/2) \ep^{abc} \om_{bc}$ [ note
the difference in the numerical factors of the quadratic terms in
(\ref{R^a}) and the bracket of the first term in the final result of
(\ref{CSgravity:comp2}) such as the latter can not be expressed as
$R^a$ only ]. The negative sign comes from $(-1)^s$ factor when we
consider $\ep_{abc} \ep^{ade}=(-1)^s (\de_b^d \de_c^e-\de_b^e
\de_c^d )$ for $s$ negative signatures in the metric $\eta_{ab}$.
This becomes (\ref{CSgravity}) in the compact form notation with the
trace $Tr$, up to the boundary term--actually this becomes a
``half'' of the Gibbons-Hawking's boundary term $2 \oint_{\cal M}
K$, for the extrinsic curvature scalar $K$ of the boundary, in the
gravity action \cite{Bana:98b}.
Note also that there are
factor ``2'' difference in the triple wedge products of $\om$'s
between (\ref{CSgravity}) and (\ref{CSgravity:comp2}).

Now, in order to determine the coefficients $\al, \be$, I need to
compare the result (\ref{CSgravity:comp2}) in the $vector$ basis
with that of the usual {\it tensor} form basis. To this end, I first
note that
\begin{\eq}
I_1 \equiv \int 2 e^a \we R_a &=& \int \ep_{abc} e^a \we R^{bc} \no \\
&=& \int \ep_{abc} e^a_{\mu} \cdot \f{1}{2} ~{R^{bc}}_{\nu \rho}
dx^{\mu}
\we dx^{\n} \we dx^{\rho}  \no \\
&=& \f{1}{2} \int d^3 x~\ep^{\m \n \rho} \ep_{abc} {e^a}_{\mu}
{R^{bc}}_{\nu \rho}   \no \\
&=& \f{1}{2} \int d^3 x~\ep^{\m \n \rho} \ep_{abc} {e^a}_{\mu}
{e^a}_{\al} {e^a}_{\be} {R^{\al \be}}_{\nu \rho}   \no \\
&=&\f{1}{2} \int d^3 x \sqrt{-g}~\ep^{\m \n \rho} \ep_{\al \be \mu}
 {R^{\al \be}}_{\nu \rho}   \no \\
 &=&-\int d^3 x \sqrt{-g}~R,
\end{\eq}
where I have denoted $R_{bc\nu \rho}=\pa_{\nu}\omega_{bc
\nu}+{\om^b}_{d \nu} \om^{dc\rho}-(\n \lra \rho)$ in the second line
and I have used $dx^{\mu} \we dx^{\nu} \we dx^{\rho}=\ep^{\m \n
\rho} ~d^3 x$ in the third line; $\ep_{abc} {e^a}_{\mu} {e^a}_{\al}
{e^a}_{\be}=e \ep_{\m \al \be}$ with $e=\sqrt{-g}$  [ $e$ is the
determinant of ${e^a}_{\mu}$ ] due to $g_{\m \n}={e^a}_{\m}
\eta_{ab} {e^b}_{\n}$ in the fourth line; the negative sign in the
final line comes from $(-1)^s$ factor with $s=1$. This is the usual
Einstein-Hilbert action, up to the sign.

Similarly, one can show that
\begin{\eq}
I_2 \equiv \int \f{1}{3 l^2}~ \ep_{abc} e^a \we e^b \we e^b &=&\int
d^3 x~\f{1}{3 l^2} ~\ep^{\m \n \rho} \ep_{abc} {e^a}_{\mu}
\we {e^b}_{\nu} \we {e^b}_{ \rho}   \no \\
&=&\int d^3 x \sqrt{-g}~\f{1}{3 l^2}~ \ep^{\m \n \rho} \ep_{\m \n
\rho} \no \\
 &=&-\int d^3 x \sqrt{-g}~\f{2}{ l^2},
\end{\eq}
where I have used $\ep^{\m \n \rho} \ep_{\m \n \rho}=(-1)^s 3!$ in
the final line. This is the cosmological constant action.

Next, I note that
\begin{\eq}
I_3 &\equiv& \int \omega^a \wedge \left( d
\omega_a+\frac{1}{3}~\ep_{abc} \omega^b \wedge \omega^c \right) \no \\
&=& \int \f{1}{2}~ \ep^{abc} \om_{bc} \wedge \left[d \left(\f{1}{2}~
\ep_{ade} \om^{de} \right)+\frac{1}{3}~\ep_{abc} \left(\f{1}{2}~
\ep^{bde} \om_{de} \right) \wedge
\left(\f{1}{2}~ \ep^{cfg} \om_{fg} \right) \right] \no \\
&=& \int \f{1}{2} \left( \om_{bc} \wedge d \om^{cb} +\frac{2}{3}
~{\om^b}_{c}  \wedge  {\om^c}_{d} \we {\om^d}_b \right).
\end{\eq}
The final line is the gravitational Chern-Simons 3-form in the
tensor basis appeared in Refs. \cite{Garc:03,Krau:05a,Saho:06} and
the first line is in the vector form basis that appeared in Refs.
\cite{Witt:88,Horn:89,Cho:98,Miel:03,Blag:03,Blag:04}, up to overall
coefficients. The relation to the component (tensor) form for the
spacetime indices is given by
\begin{\eq}
I_3 &=& \int \f{1}{2} \left( \om_{bc} \wedge R^{cb} -\frac{1}{3}
~{\om^b}_{c}  \wedge  {\om^c}_{d} \we {\om^d}_b \right) \no \\
&=& \int \f{1}{2} \left( \om_{bc \m}\cdot \f{1}{2} {R^{cb}}_{\n
\rho} -\frac{1}{3} ~{\om^b}_{c \m}   {\om^c}_{d \n} {\om^d}_{b \rho}
\right) dx^{\m} \we
dx^{\n} \we dx^{\rho} \no \\
&=& -\int d^3 x ~\f{1}{4}~ \ep^{\m \n \rho} ~\left( \om_{bc \m}
{R^{bc}}_{\n \rho} +\frac{2}{3} ~{\om^b}_{c \m} {\om^c}_{d \n}
{\om^d}_{b \rho} \right).
\end{\eq}
This expression is what appeared in Refs.
\cite{Dese:82,Dese:91,
Solo:05a}.

Finally, I note that
\begin{\eq}
I_4 \equiv \int \frac{1}{l^2} e^a \wedge { T_a} &=& \int
\frac{1}{l^2} {e^a}_{\m}\cdot \f{1}{2} T_{a\n \rho} dx^{\m} \we
dx^{\n} \we dx^{\rho} \no \\
&=& \int d^3x~\frac{1}{l^2} ~\ep^{\m \n \rho}~{e^a}_{\m}  T_{a\n
\rho} dx^{\m} \we dx^{\n} \we dx^{\rho},
\end{\eq}
where $ T_{a\n \rho}=\pa_{\n} {e^a}_{\rho}+ {\ep^a}_{bc}
{\om^b}_{\n} {e^c}_{\rho}-(\n \lra \rho)$ is the torsion tensor.
This action is what appeared in Refs.
\cite{Witt:88,Cho:98,Blag:03,Blag:04}.

Collecting all formulae together, I arrive at the following action
for the generalized CS gravity, up to the boundary term in
(\ref{CSgravity:comp2}),
\begin{\eq}
I_{CS}&=&\f{k}{4 \pi}~ \al (I_3+I_4) +\f{k}{4 \pi}~ \be (I_1 +I_2)
\no
\\
&=&-\f{k \be}{4 \pi} \int_{\cal M} d^3 x \sqrt{-g}~\left( R
+\f{2}{l^2} \right) -\f{k \al}{16 \pi} \int_{\cal M} d^3 x ~ \ep^{\m
\n \rho} ~\left( \om_{bc \m} {R^{cb}}_{\n \rho} +\frac{2}{3}
~{\om^b}_{c \m} {\om^c}_{d \n}
{\om^d}_{b \rho} \right) \no \\
&&+ \f{k \al}{8 \pi} \int_{\cal M} d^3x~\frac{1}{l^2} ~\ep^{\m \n
\rho}~{e^a}_{\m} T_{a\n \rho}.
\end{\eq}
This is the expression that appeared in Refs.
\cite{Dese:82,Dese:91,
Solo:05a}, but it is easy to compare
with other expressions in Refs.
\cite{Krau:05a,Saho:06,Blag:03,Blag:04} from the above formulae.
Now, in order that the first term becomes the ordinary
Einstein-Hilbert action $I_{EH}=(1/16 \pi G) \int_{\cal M} (R
+2/l^2)$ with a negative cosmological constant $\La=-1/l^2$ in
(\ref{EHGCS}) I choose $k \be=-1/4G$, as in (\ref{connect}). Then,
the gravitational Chern-Simons term becomes, in several equivalent
expressions,
\begin{\eq}
I_{GCS} & \equiv & \f{k}{4 \pi}~ \al I_3 \no \\
&=&\f{1}{64 \pi G} \f{\al}{\be} \int_{\cal M} d^3 x ~ \ep^{\m \n
\rho} ~\left(  \om_{bc \m} {R^{bc}}_{\n \rho} +\frac{2}{3}
~{\om^b}_{c \m} {\om^c}_{d \n}
{\om^d}_{b \rho}\right) \no \\
&=&-\f{1}{32 \pi G} \f{\al}{\be} \int_{\cal M} \left( \om_{bc}\we d
\om^{cb} +\frac{2}{3} ~{\om^b}_{c \m} {\om^c}_{d \n}
{\om^d}_{b \rho} \right) \no \\
&=&-\f{1}{16 \pi G} \f{\al}{\be} \int_{\cal M} \om^{a}\we \left(  d
\om_{a} +\frac{1}{3}~\ep_{abc}  \om^b \we \om^c \right).
\end{\eq}
By comparing the first line with (\ref{GCS}) and Refs.
\cite{Dese:82,Dese:91},
\cite{Solo:05a} (the published version), I find $\hb=\al/l
\be=-1/\mu l=-\be_S/l$ for the coefficient $\mu$ in Refs.
\cite{Dese:82,Dese:91}
and $\be_S$ in Ref. \cite{Solo:05a},
as I have claimed in (\ref{connect}); by comparing the second line
with Ref. \cite{Krau:05a}, I find $\hb=\al/l \be=-32 \pi G
\be_{KL}/l$ for the coefficient $\be_{KL}$ in Ref. \cite{Krau:05a};
by comparing the third line with Refs. \cite{Blag:03,Blag:04}, I
find $\hb=\al/l \be=-16 \pi G \al_3/l$. From these relations one can
ensure that the central charges between the anomaly approaches of
Refs. \cite{Krau:05a,Solo:05a} and the classical symmetry approaches
of Refs. \cite{Blag:03,Park:06} agree exactly, even in the presence
of gravitational Chern-Simons term,
\begin{\eq}
\hat{c}^{\pm}_{\tot}&=&\f{1}{\hbar} \left(1 \mp 16 \pi G
\f{\al_3}{l}
\right) \f{3l}{2 G} \no \\
&=&\f{1}{\hbar} \left(1 \mp 32 \pi G  \f{\be_{KL}}{l}
\right) \f{3l}{2 G} \no \\
&=&\f{1}{\hbar} \left(1 \mp \f{\be_{S}}{l}
\right) \f{3l}{2 G} \no \\
&=&\f{1}{\hbar} \left(1 \pm \hb \right) \f{3l}{2 G}.
\end{\eq}
\\

\begin{section}
{Cardy formula and its higher-order corrections}
\end{section}

In this appendix, I briefly review the physicist's  derivation of
Cardy formula and its higher-order corrections, for completeness of
my discussions in this paper.

To this end, let me begin with the partition function of the CFT on
a torus, with the modular parameters $\tau, \bar{\tau}$
\ci{Card:86,Park:04b}
\begin{\eq}
\label{CFTZ}
 Z[\tau,\bar{\tau}]=Tr e^{2 \pi i \tau (\hat{L}_0-\f{\hat{c}}{24})}
e^{-2 \pi i \bar{\tau} (\hat{\bar{L}}_0-\f{\hat{\bar{c}}}{24})}.
\end{\eq}
This is invariant under the modular transformations $\tau\ra (a \tau
+b)/(c \tau +d)$ (similarly for $\bar{\tau}$), with the some
integers $a,b,c,d$ satisfying $ad-bc=1$, and the Virasoro generators
$\hat{L}_m,\hat{\bar{L}}_m$ are defined on the ``plane'' with
central charges $\hat{c}, \hat{\bar{c}}$, with the algebras in the
standard form,
\begin{\eq}
&&[\hat{L}_m , \hat{L}_n]=(m-n)\hat{L}_{m+n} +\f{\hat{c}}{12} m (m^2-1) \de_{m+n,0}, \no \\
&&[\hat{\bar{L}}_m, \hat{\bar{L}}_n]=(m-n) \hat{\bar{L}}_{m+n} +\f{\hat{\bar{c}}}{12} m (m^2-1) \de_{m+n,0}, \no \\
&&[\hat{L}_m, {\hat{\bar L}}_n]=0.
\end{\eq}

The density of states $\rho(\hat{\De}, \hat{\bar{\De}})$ for the
eigenvalues $\hat{L}_0=\hat{\De}, \hat{\bar{L}}_0=\hat{\bar{\De}}$
is given as a contour integral (I suppress the
$\bar{\tau}$-dependence for simplicity, but the computation is
similar to the $\tau$-part)
\begin{\eq}
\label{rho}
 \rho(\hat{\De}) =\int_C d\tau~ e^{-2\pi i (\hat{\De} -\f{\hat{c}}{24})
\tau } Z[\tau],
\end{\eq}
where the contour $C$ encircles the origin in the complex $q=e^{2
\pi i \tau}$ plane.
The general evaluation of this integral would be impossible unless
$Z[\tau]$ is known completely. But, due to the modular invariance of
(\ref{CFTZ}), one can easily compute its asymptotic formula through
the steepest-descent approximation. In particular, (\ref{CFTZ}) is
invariant under $\tau \ra -1/\tau$ \ci{Card:86} such that
\begin{\eq}
Z[\tau] =Z[-1/\tau]=e^{-2 \pi i (\hat{\De}_{\hbox{\scriptsize
min}}-\f{\hat{c}}{24}) \tau }\tilde{Z}[-1/\tau],
\end{\eq}
where $\tilde{Z}[-1/\tau]=Tr e^{-2 \pi i
(\hat{L}_0-\hat{\De}_{\hbox{\scriptsize min}})/ \tau}$ approaches a
constant value $\rho(\hat{\De}_{\hbox{\scriptsize min}})$ as $\tau
\ra i 0_+$, which defines the steepest-descent path for a ``real''
value of $\hat{\De} \geq \hat{\De}_{\hbox{\scriptsize min}}$. With
the help of this property, (\ref{rho}) is evaluated as, by expanding
the integrand
around the steepest-descent path $\tau_*$,
\begin{\eq}
\label{rho-ser}
 \rho(\hat{\De}) &=&\int_C d\tau~ e^{\eta(\tau)} \tilde{Z}[-1/\tau] \\
&=&e^{\eta(\tau_*)}  \tilde{Z}[-1/\tau^*] \times \int_C d \tau
~\mbox{exp} \left\{ \f{1}{2} \eta^{(2)} (\tau-\tau_*)^2
+\sum^{\infty}_{n\geq
3} \f{1}{n !} \eta^{(n)} (\tau-\tau_* )^n \right\} \no \\
 &&\times \left[ 1+\sum^{\infty}_{m \geq 1} \f{1}{m !} \tilde{Z}^{-1}
\tilde{Z}^{(m)} (\tau-\tau_*)^m \right]. \label{rho-expand}
\end{\eq}
Here, $\eta(\tau)=-2 \pi i \hat{\De}_{\eff} \tau +{2 \pi i}
\hat{c}_{\eff}/({24}{\tau})$, which dominates $\tilde{Z}[1/\tau]$ in
the region of interest, gets the maximum
\begin{\eq}
\eta(\tau_*) &=&2 \pi \sqrt{\f{\hat{c}_{\eff} \hat{\De}_{\eff}}{6}},
\end{\eq}
with
$
\tau_*=i \sqrt{{\hat{c}_{\eff}}/{24 \hat{\De}_{\eff}}},
$
when
\begin{\eq}
\label{cond} \f{24 \hat{\De}_{\eff}}{\hat{c}_{\eff}} \gg 1
\end{\eq}
is satisfied. Here, $\eta^{(n)}= ({d^n \eta}/{d
\tau^n})|_{\tau=\tau_*}, ~\tilde{Z}^{(m)}=({d^n \tilde{Z}}/{d
\tau^n})|_{\tau=\tau_*} $, and $\hat{c}_{\eff}=\hat{c}-24
\hat{\De}_{\hbox{\scriptsize min}},~
\hat{\De}_{\eff}=\hat{\De}-\hat{c}/{24}$;
$\hat{\De}_{\hbox{\scriptsize min}}$ is the minimum of $\hat{\De}$.
Here, I am assuming ``$\hat{c}_{\eff}, \hat{\De}_{\eff}>0$'' since,
otherwise, the saddle-point approximation is not valid for $real$
valued $\hat{c}_{\eff},~\hat{\De}_{\eff}$.

Then, in the limit of $\ep \ra \infty$ with $\tau_*=i/\ep$, the
higher-order correction terms in the bracket $[~~]$ of
(\ref{rho-expand}) are exponentially suppressed as $e^{- 2 \pi \ep
(\hat{\De}-\hat{\De}_{\hbox{\scriptsize min}})}$, hence
(\ref{rho-expand}) is simplified as, up to the exponentially
suppressing terms,
\begin{\eq}
\label{rho-ser'}
 \rho(\hat{\De}) =e^{2 \pi \sqrt{\hat{c}_{\eff} \hat{\De}_{\eff}/6}}
  \times \int_C d\tau~ \mbox{exp}
\left\{ \f{1}{2} \eta^{(2)} (\tau-\tau_*)^2 +\sum^{\infty}_{n\geq 3}
\f{1}{n !} \eta^{(n)} (\tau-\tau_* )^n \right\},
\end{\eq}
where I have used $\tilde{Z} [i \infty] =1$. This is known as the
Cardy formula \ci{Card:86}. Note that here I need
\begin{\eq}
\hat{c}_{\eff} \hat{\De}_{\eff} \gg 1
\end{\eq}
 in order that the approximation is
reliable, i.e., $e^{\eta(\tau_*)}$ dominates in the integral of
(\ref{rho-ser}), as well as the condition (\ref{cond}), such as
$\tilde{Z}[-1/\tau]$ is slowly varying near $\tau_*$.

The integrals above could be evaluated by the steepest-descent
method but the direct computation would be quite involved if one
wants to go beyond the Gaussian integral. But fortunately there
exits an {\it exact}, closed expression, due to Raedmacher
\cite{Rade:38}, with the result \cite{Dijk:00,Birm:00}
\begin{\eq}
\label{rho:CFT4} \rho(\hat{\De}) =e^{2 \pi \sqrt{\hat{c}_{\eff}
\hat{\De}_{\eff}/6}} \times \left(\f{\hat{c}_{\eff}}{96
\hat{\De}_{\eff}^3}\right)^{1/4} I_1(2 \pi \sqrt{\hat{c}_{\eff}
\hat{\De}_{\eff}/6}),
\end{\eq}
up to the exponentially suppressed terms. So, its corresponding
entropy $S_{\stat}=\log \rho(\hat{\De})$ becomes, with $S_0=2 \pi
\sqrt{\hat{c}_{\eff} \hat{\De}_{\eff}/6}$,
\begin{\eq}
\label{SCFT4}
S_{\stat}&=&S_0+\mbox{ln}\left[\left(\f{\hat{c}_{\eff}}{96
\hat{\De}_{\eff}^3}\right)^{1/4}
I_1(S_0)   \right] \no \\
&=&S_0+\mbox{ln}\left(\f{\hat{c}_{\eff}}{96
\hat{\De}_{\eff}^3}\right)^{1/4}-\f{3}{8} S_0^{-1} +O( (S_0)^{-2}),
\end{\eq}
where $I_n(x)$ is the modified Bessel function of the first kind,
and I have used its asymptotic series expansion for large $x$:
\begin{\eq}
 I_1(x)=\f{1}{\sqrt{2 \pi x}} e^x\left[1-\f{3}{8
} x^{-1}+O(x^{-2})\right].
\end{\eq}

\section*{Acknowledgments}

I would like to thank Jacob Bekenstein, Jin-Ho Cho, Gungwon Kang,
O-kab Kwon, Makoto Natsuume, Sergei Odintsov, Segrey Soloduhkin, and
Ho-Ung Yee for useful correspondences. This work was supported by
the Science Research Center Program of the Korea Science and
Engineering Foundation through the Center for Quantum Spacetime
(CQUeST) of Sogang University with grant number R11 - 2005- 021.

\newcommand{\J}[4]{#1 {\bf #2} #3 (#4)}
\newcommand{\andJ}[3]{{\bf #1} (#2) #3}
\newcommand{\AP}{Ann. Phys. (N.Y.)}
\newcommand{\MPL}{Mod. Phys. Lett.}
\newcommand{\NP}{Nucl. Phys.}
\newcommand{\PL}{Phys. Lett. }
\newcommand{\PR}{Phys. Rev. }
\newcommand{\PRL}{Phys. Rev. Lett.}
\newcommand{\PTP}{Prog. Theor. Phys.}
\newcommand{\CQG}{Class. Quant, Grav.}
\newcommand{\hep}[1]{ hep-th/{#1}}
\newcommand{\hepg}[1]{ gr-qc/{#1}}
\newcommand{\bi}{ \bibitem}

\end{document}